\documentclass[apj, numberedappendix]{emulateapj}

\newcommand{\lum}{luminosity}
\newcommand{\lums}{luminosities}

\newcommand{\spitzer}{\textit{Spitzer}}

\newcommand{\Mstar}{\hbox{$M_*$}}
\newcommand{\Msol}{\hbox{$M_\odot$}}

\newcommand{\msol}{\hbox{$M_\odot$}}

\newcommand{\lsol}{\hbox{$L_\odot$}}

\newcommand{\uJy}{\hbox{$\mu$Jy}}

\newcommand{\um}{\hbox{$\mu$m}}

\newcommand{\mone}{\hbox{$[3.6]$}}
\newcommand{\mtwo}{\hbox{$[4.5]$}}
\newcommand{\mthree}{\hbox{$[5.8]$}}
\newcommand{\mfour}{\hbox{$[8.0]$}}
\newcommand{\degree}{\ensuremath{^\circ}}%
\newenvironment{my_itemize}{
\begin{itemize}
  \setlength{\itemsep}{1pt}
  \setlength{\parskip}{0pt}
  \setlength{\parsep}{0pt}}{\end{itemize}
}

\slugcomment{Accepted for publication in the Astrophysical Journal}

\shorttitle{the SIMPLE survey}
\shortauthors{Damen et al.}

\def\figa{
  \begin{figure*}[p]  
    \includegraphics[width=\textwidth]{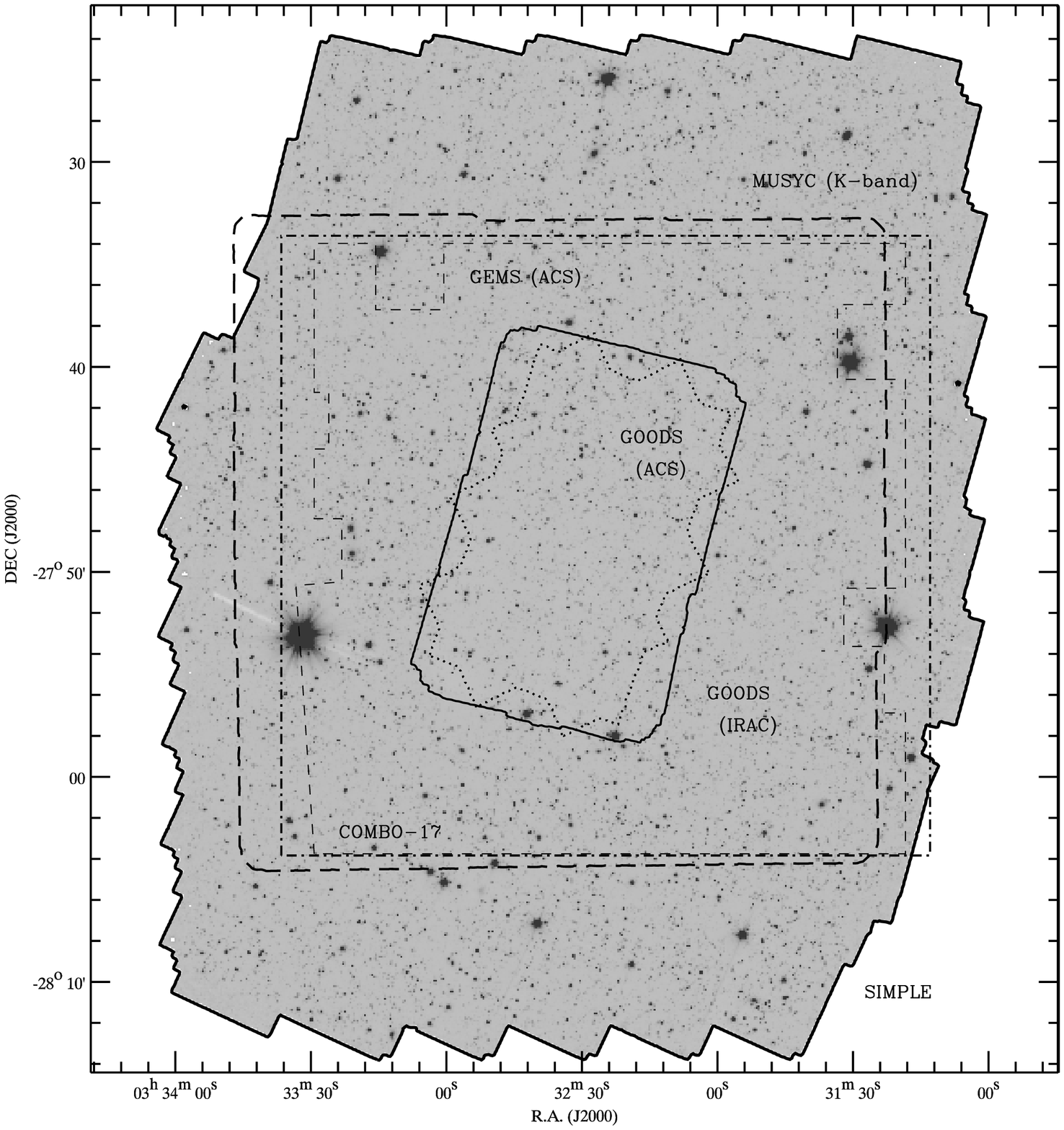}
    \caption[fov.ps]{E-CDFS in the combined 3.6 \um\,+ 4.5 \um\,
      detection image. The image is normalized by the square root of
      the weight map, producing a noise-equalized detection image (see
      Section \ref{det}). The thin dashed lines delineate the GEMS
      field, COMBO-17 is represented by the dash-dotted lines, the
      dotted and solid lines indicate the field of view of the GOODS
      ACS and IRAC observations, respectively, and the long dashed
      lines indicate the MUSYC field.
      \label{fov}}  
  \end{figure*}
}

\def\figb{
  \begin {figure}[b] 
    \centering
    \includegraphics[width=0.4\textwidth]{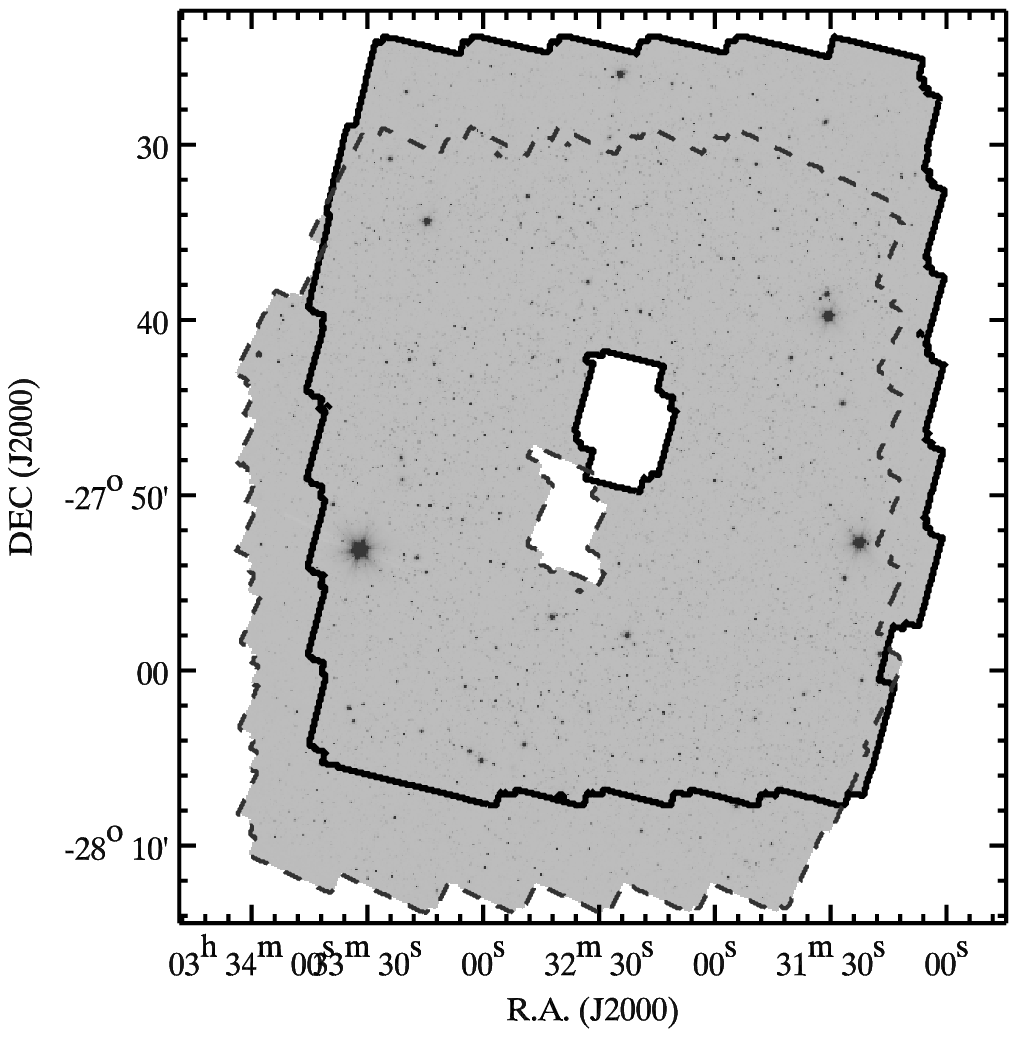}
    \hspace{0.4in}
    \includegraphics[width=0.4\textwidth]{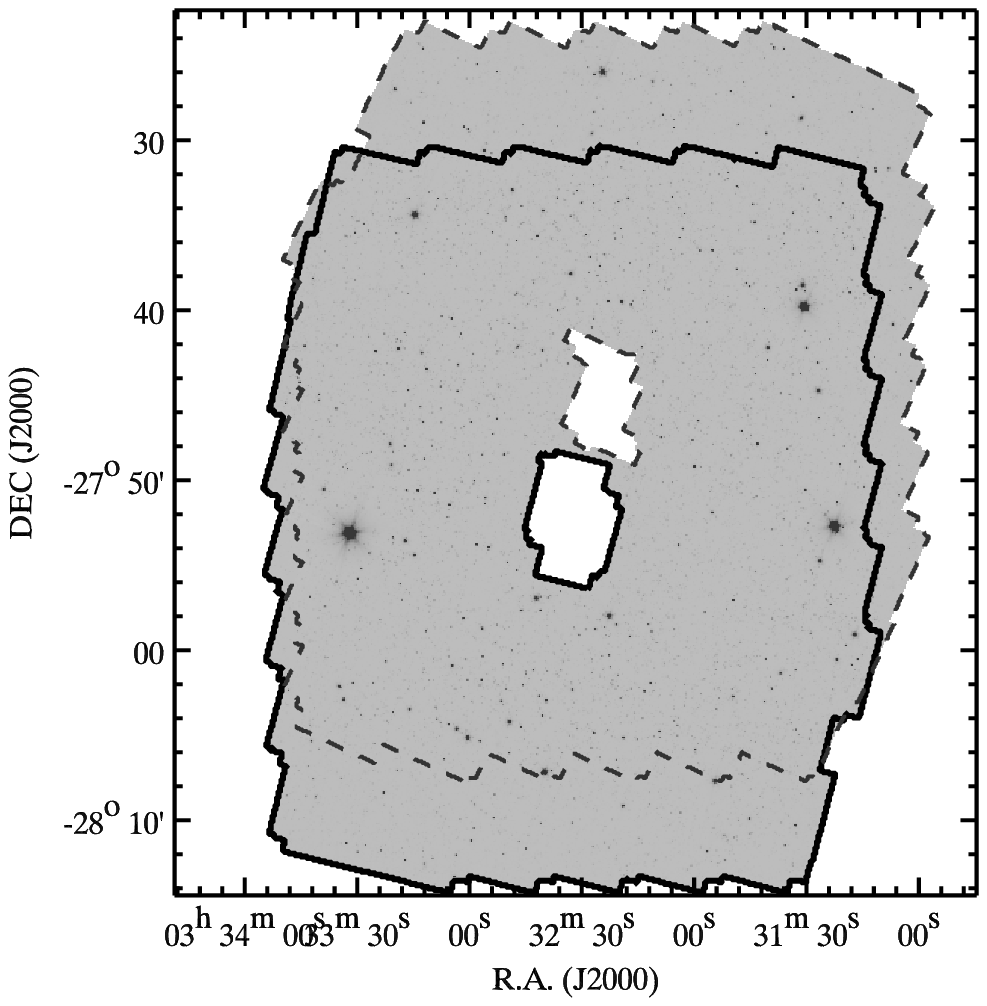}
    \caption{E-CDFS in channel 1 ({\it upper panel}) and channel 2 ({\it
        lower panel}). In both panels the data of the first epoch are
      indicated by the solid lines and those of the second epoch with
      dashed lines. Due to the special setup of IRAC, the full area is
      covered after the two epochs for all channels. Since channels 1
      and 3 are observed simultaneously, the lines in the left panel
      also delineate the field of view of channel 3. The same is true
      for channels 2 and 4 in the right panel.
      \label {fov_ep}
    }
  \end {figure}
}

\def\figc{
\begin{figure*}
    \centering
    \includegraphics[width=0.3\textwidth]{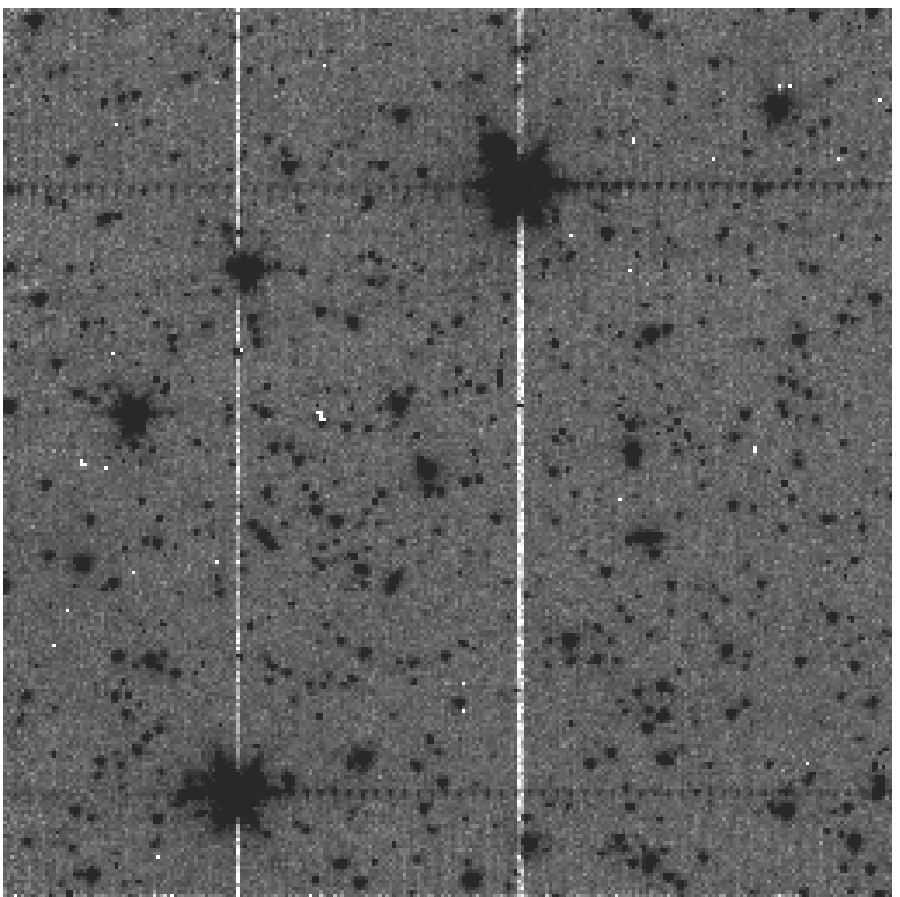}
    \hspace{0.2in}
    \includegraphics[width=0.3\textwidth]{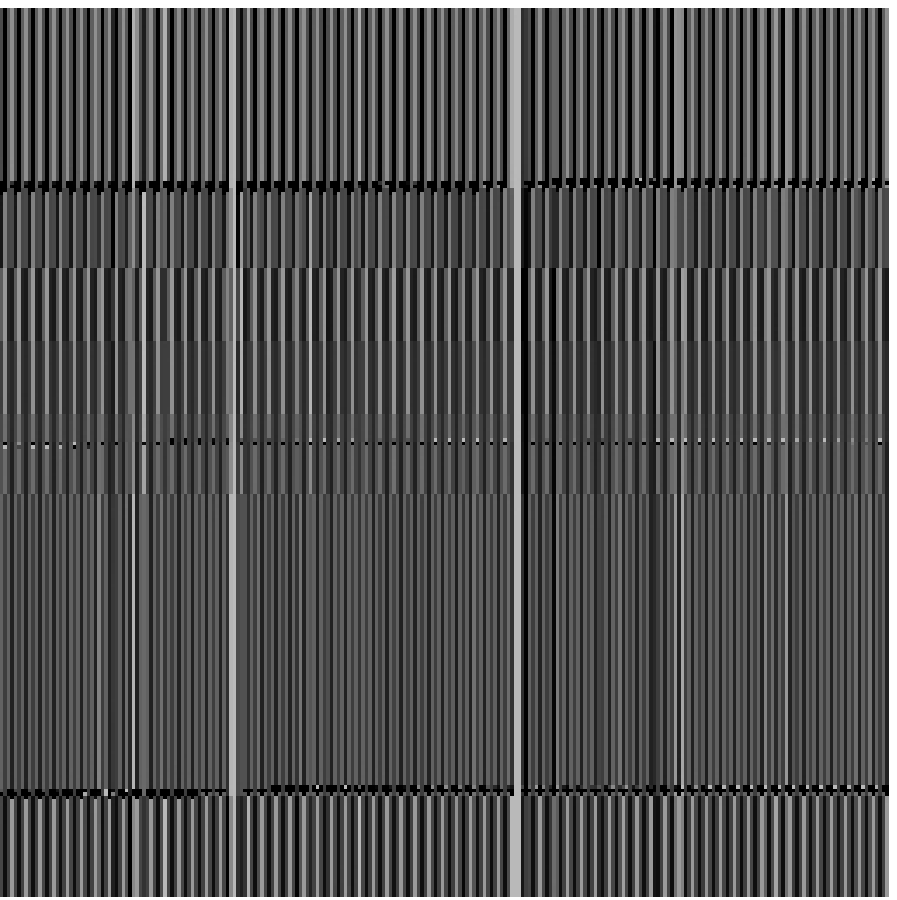}
    \hspace{0.2in}
    \includegraphics[width=0.3\textwidth]{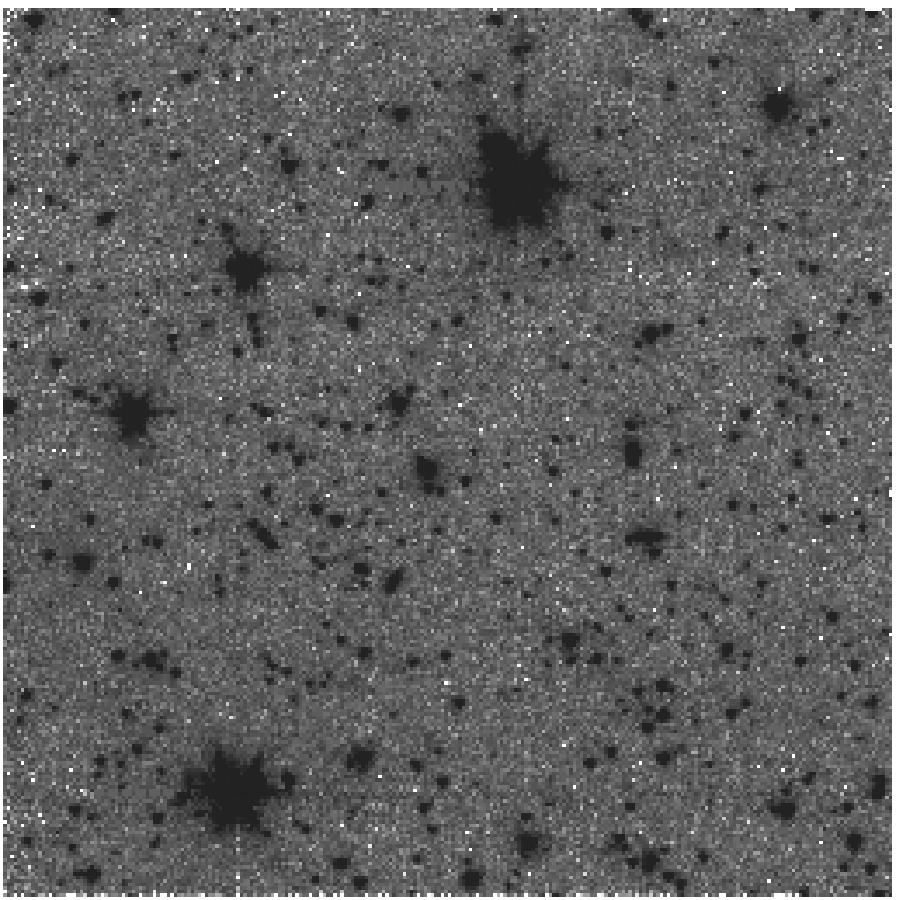}
    \caption[see.ps]{{\it Left} - Typical BCD frame, suffering from
      muxbleed (the horizontal black pattern of both sides of the
      bright sources), column pulldown (vertical white lines), and
      muxstriping (jailbar pattern that extends below each bright
      sources over the full width of the frame). {\it Center} -
      Correction image that is subtracted from the affected
      frame. {\it Right} - Cleaned image, after subtraction of the
      center frame and removal of cosmic rays. Image from Astronomical
      Observation Request (AOR) r15564288, channel 1, 96.4 s
      exposure time. 
      \label{art1}}
  \end{figure*}
}

\def\figd{
    \begin{figure*}
    \centering
    \includegraphics[width=0.3\textwidth]{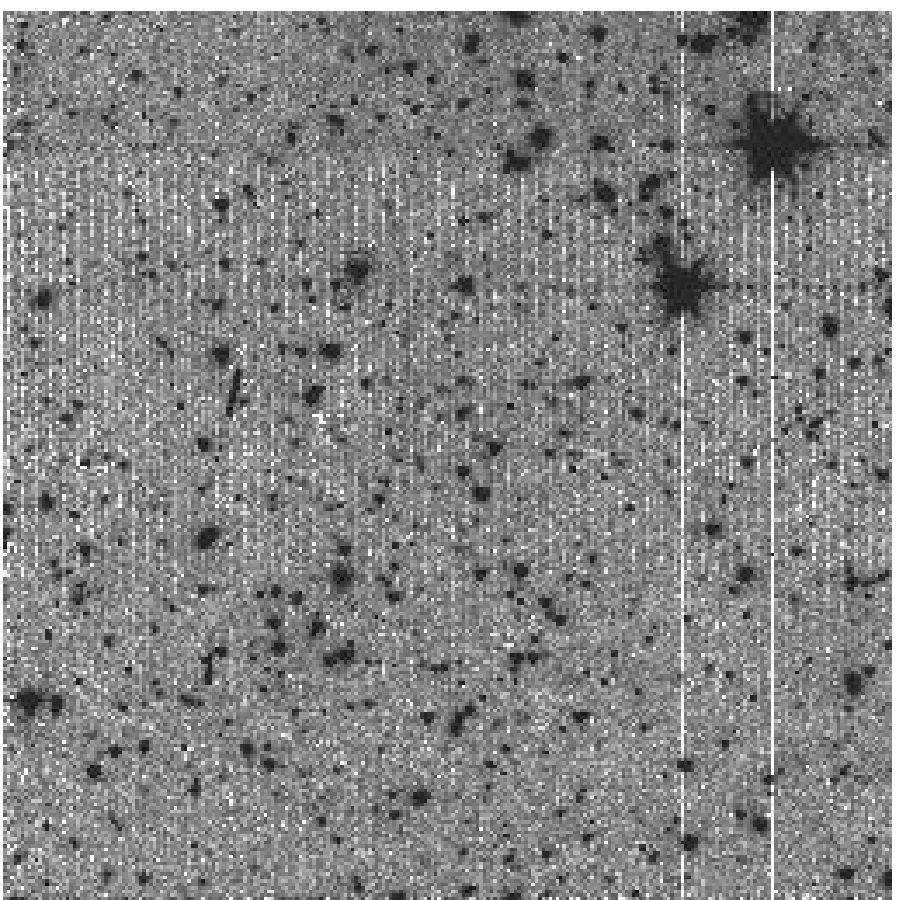}
    \hspace{0.2in}
    \includegraphics[width=0.3\textwidth]{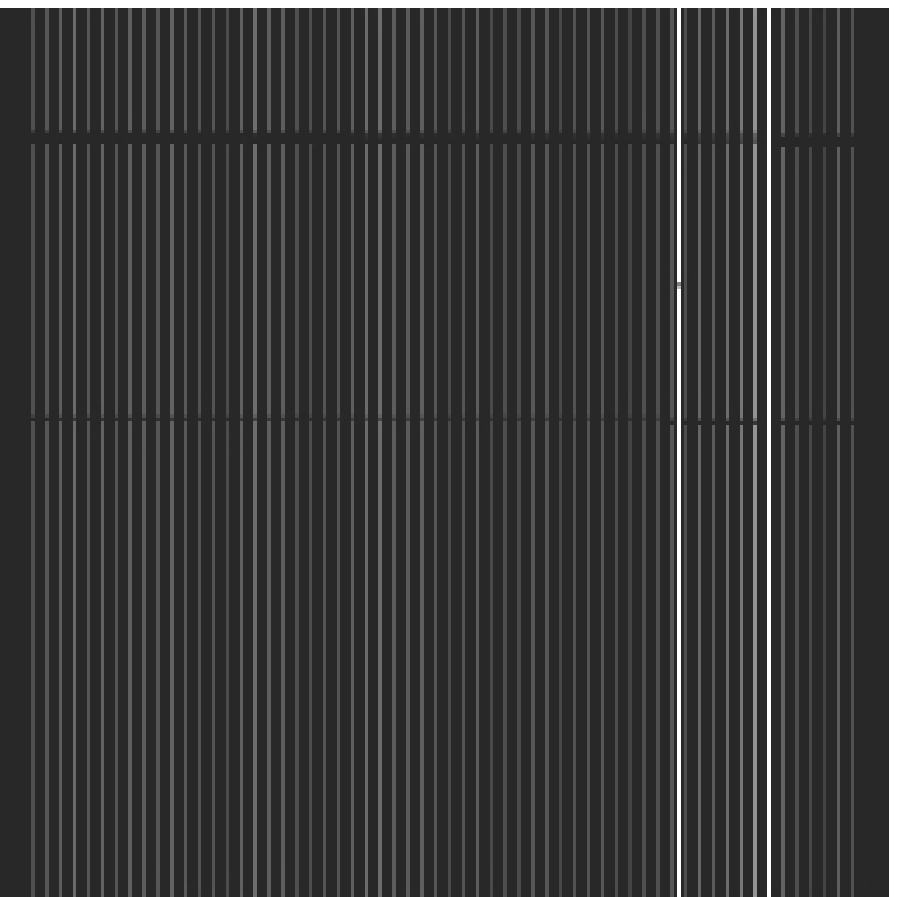}
    \hspace{0.2in}
    \includegraphics[width=0.3\textwidth]{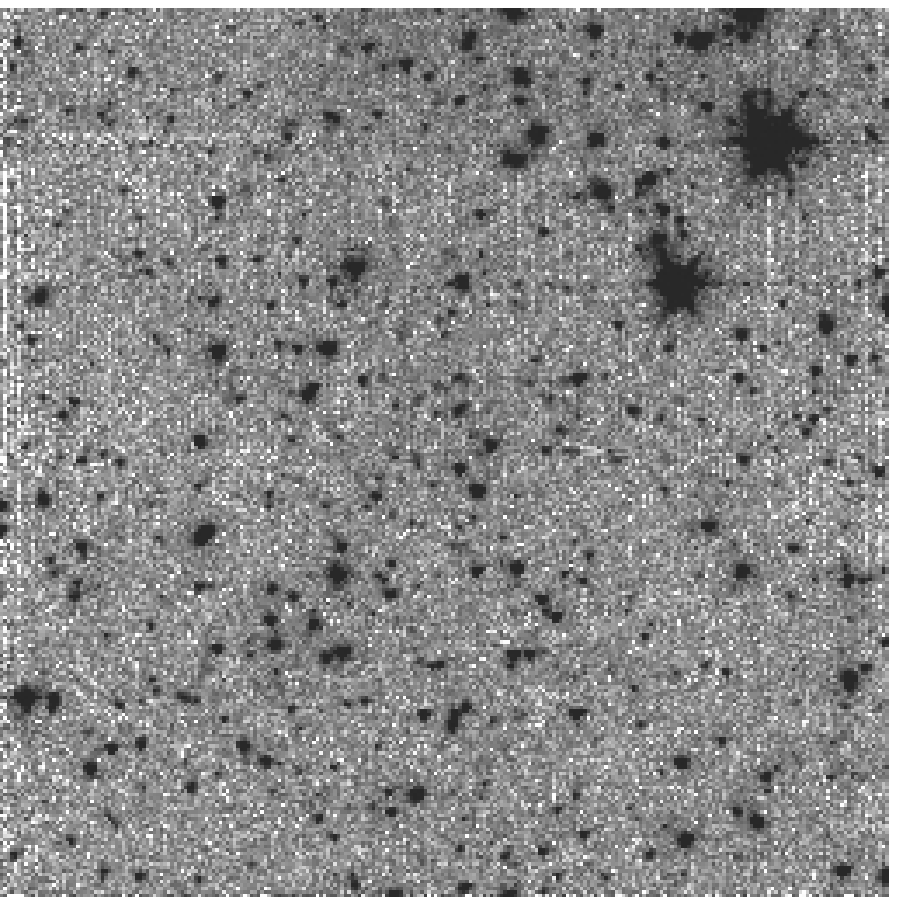}
    \caption[see.ps]{Same as Fig. \ref{art1}, with more pronounced
      muxstriping pattern. Image from AOR r15564032, channel 1, 96.4
      s exposure time.
      \label{art2}}
  \end{figure*}
}

\def\fige{
  \begin{figure}
    \centering
    \includegraphics[width=0.5\textwidth]{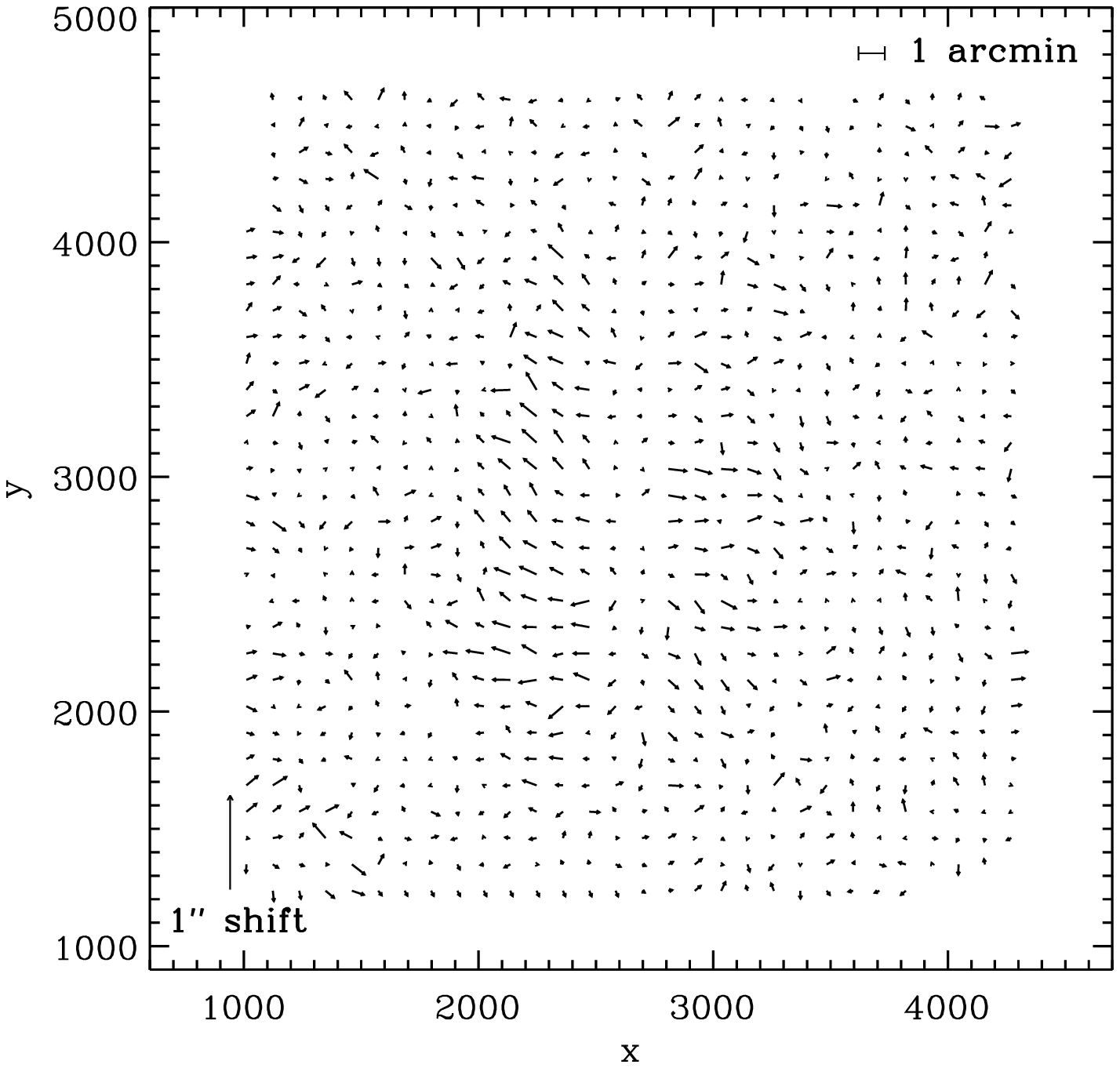}
    \caption[see.ps]{Map of residual shifts of compact sources in the
      3.6 \um\, image with respect  to a compact-source catalog
      detected in the deep $BVR$ image. Large-scale shears, systematic
      variations on scales of a few arcminutes, are 0$\arcsec$.2 or
      less.
      \label{astro}}
  \end{figure}
  }

\def\figf{
\begin{figure*}[p] 
    \includegraphics[width=\textwidth]{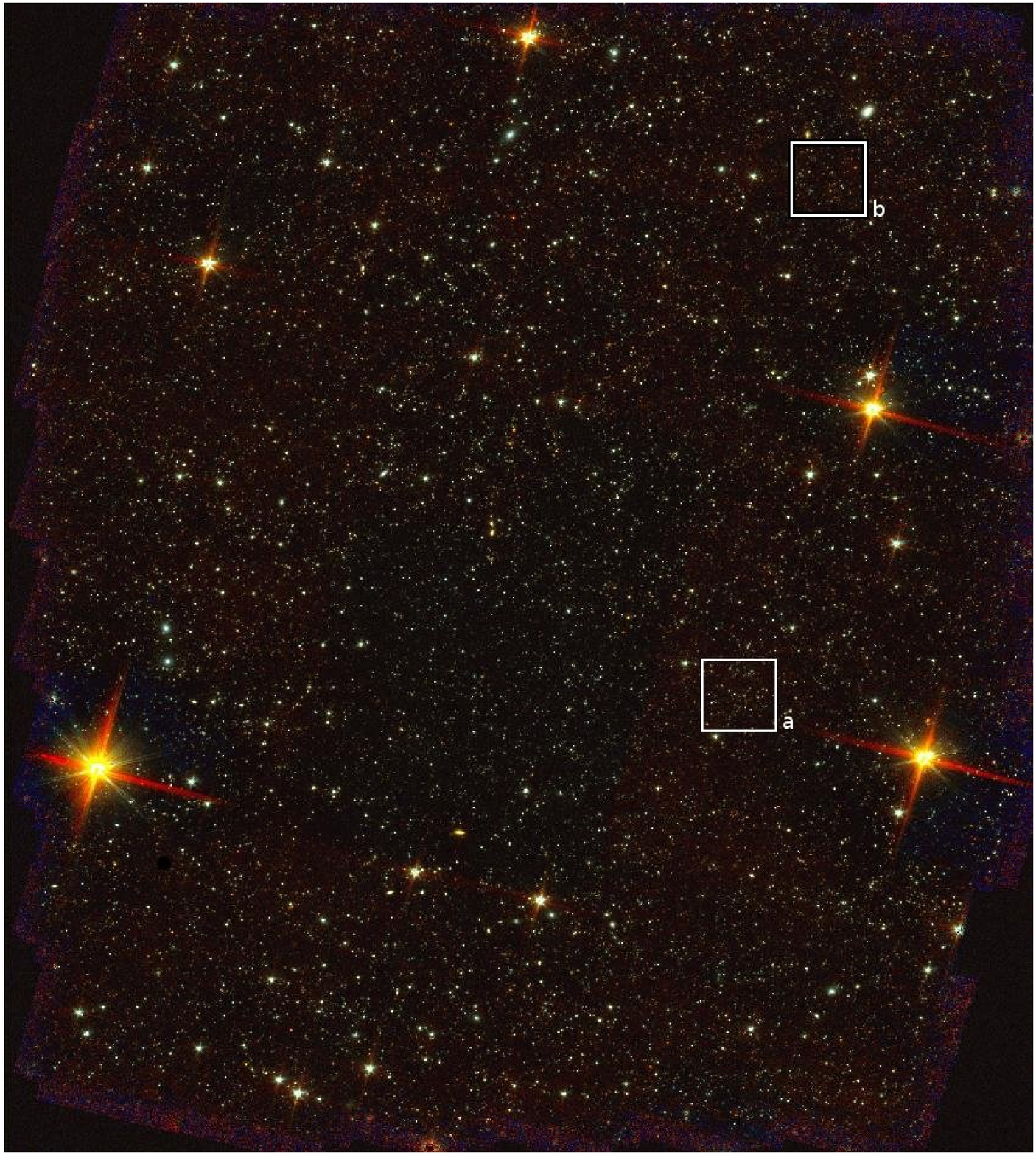}
    \caption[col.jpg]{Two-color composite image of the IRAC data of
      the E-CDFS, based on the 3.6 \um\, and 5.8 \um\, bands. The
      total field size is 38\arcmin \, $\times$ 48\arcmin\, and north is
      up. Figure~\ref{colorz} shows zoomed-in versions of the areas
      outlined in white.
      \label{color}}
  \end{figure*}
}

\def\figg{
  \begin {figure}[b!]
    \centering
    \includegraphics[width=0.4\textwidth]{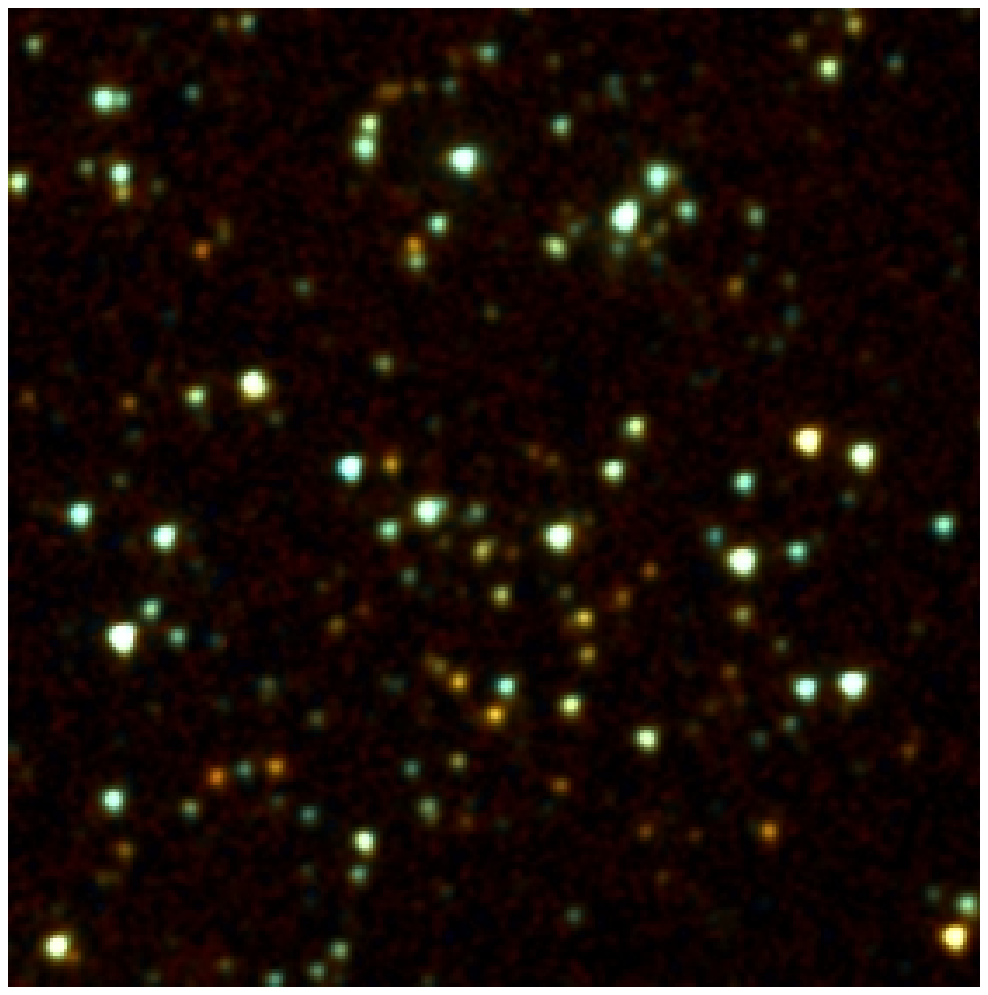}
    \hspace{0.4in}
    \includegraphics[width=0.4\textwidth]{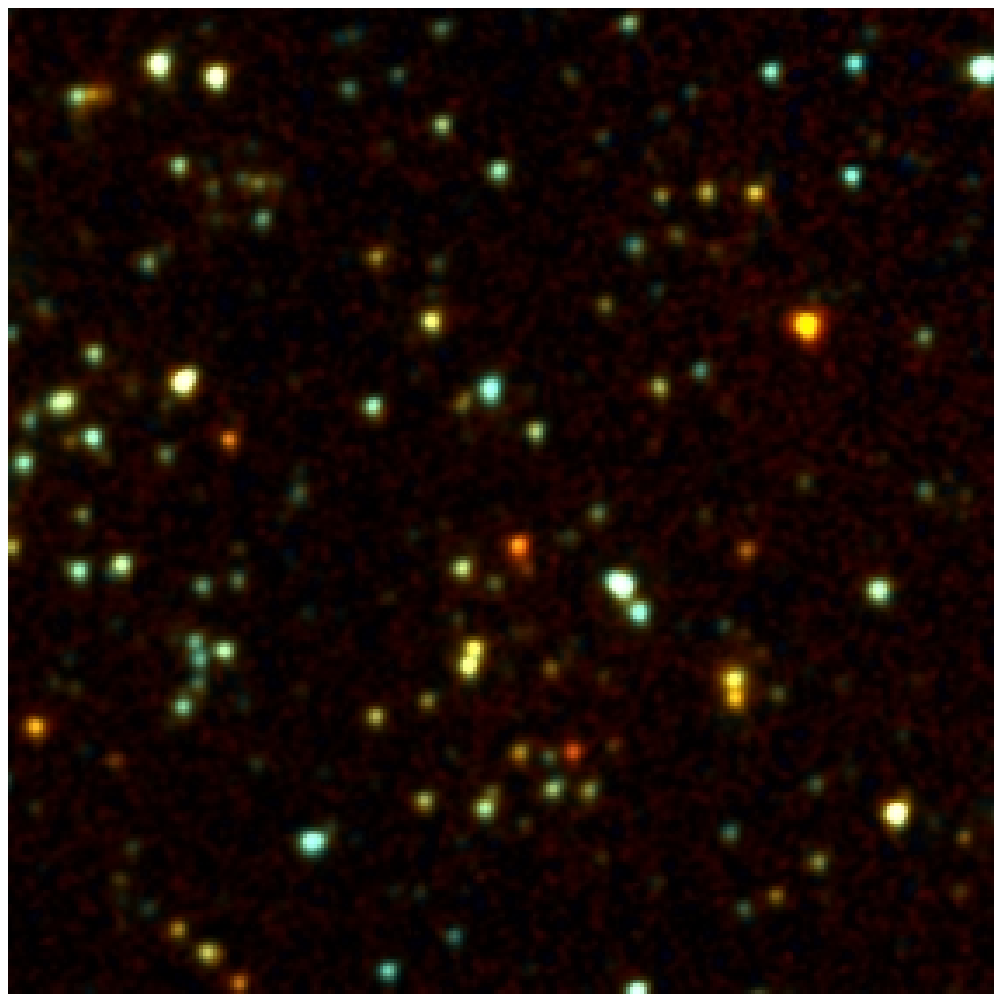}
    \caption{Two example zoomed-in cut-out areas showing details of
      the full mosaic indicated in Fig.~\ref{color}: (a) left, and (b)
      right. The images have been enlarged 20 times. The field
      size is 2\arcmin.5 \, $\times$ 2\arcmin.5.
      \label {colorz}
    }
  \end {figure}
}

\def\figh{
  \begin{figure*}
    \includegraphics[width=\textwidth]{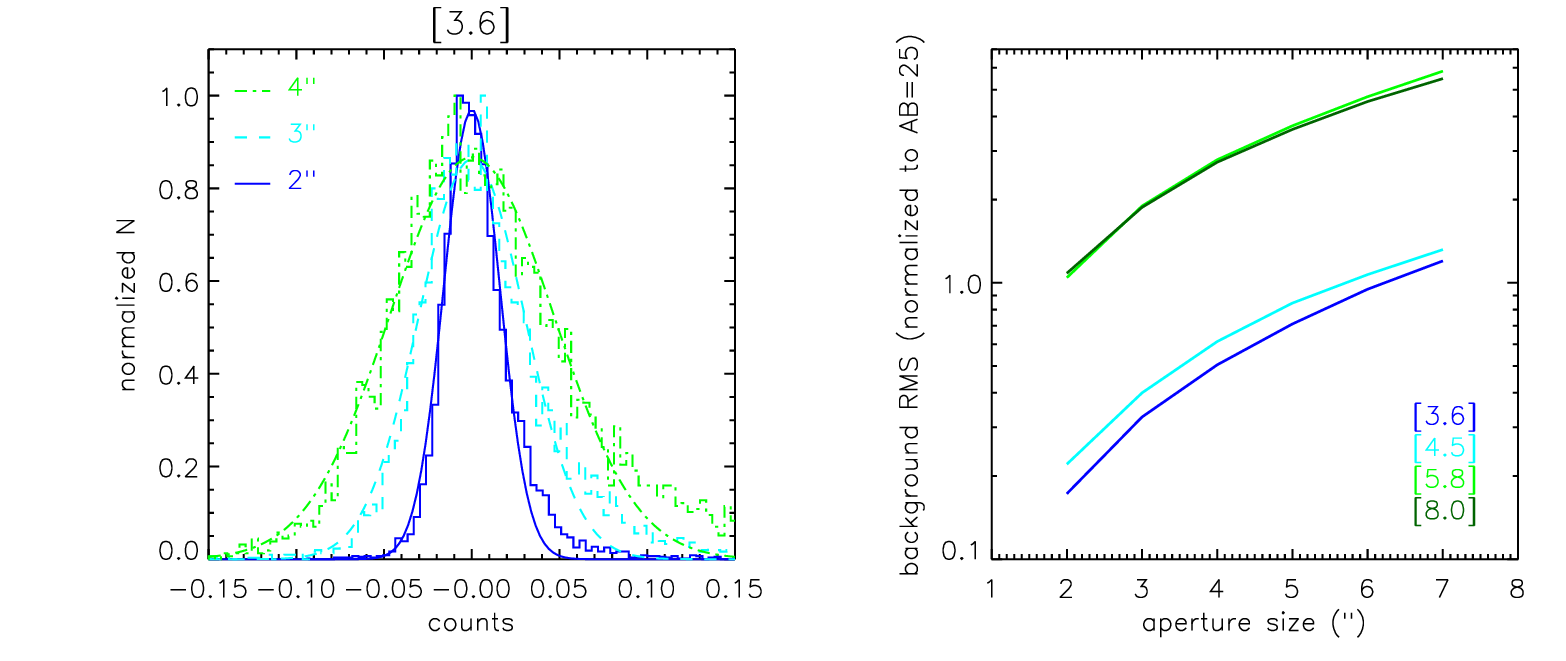}
    \caption[rms]{Background RMS derived from the distributions of
      fluxes within randomly placed empty apertures. {\it Left} -
      Distribution of empty aperture fluxes within a 2" ({\it blue}),
      3" ({\it cyan}), and 4" ({\it green}) aperture diameter on the
      IRAC 3.6 \um\, image. The distribution is well described by a
      Gaussian with an increasing width for increasing aperture size. 
      {\it Right} - Background RMS as derived from flux measurements
      within empty apertures versus aperture size for the IRAC bands
      3.6 \um\,({\it blue}), 4.5 \um\,({\it cyan}), 5.8 \um\,({\it
        green}) and 8.0 \um\,({\it dark green}). 
      \label{rms}}
  \end{figure*}
}

\def\figi{
 \begin{figure}
   \includegraphics[width=0.5\textwidth]{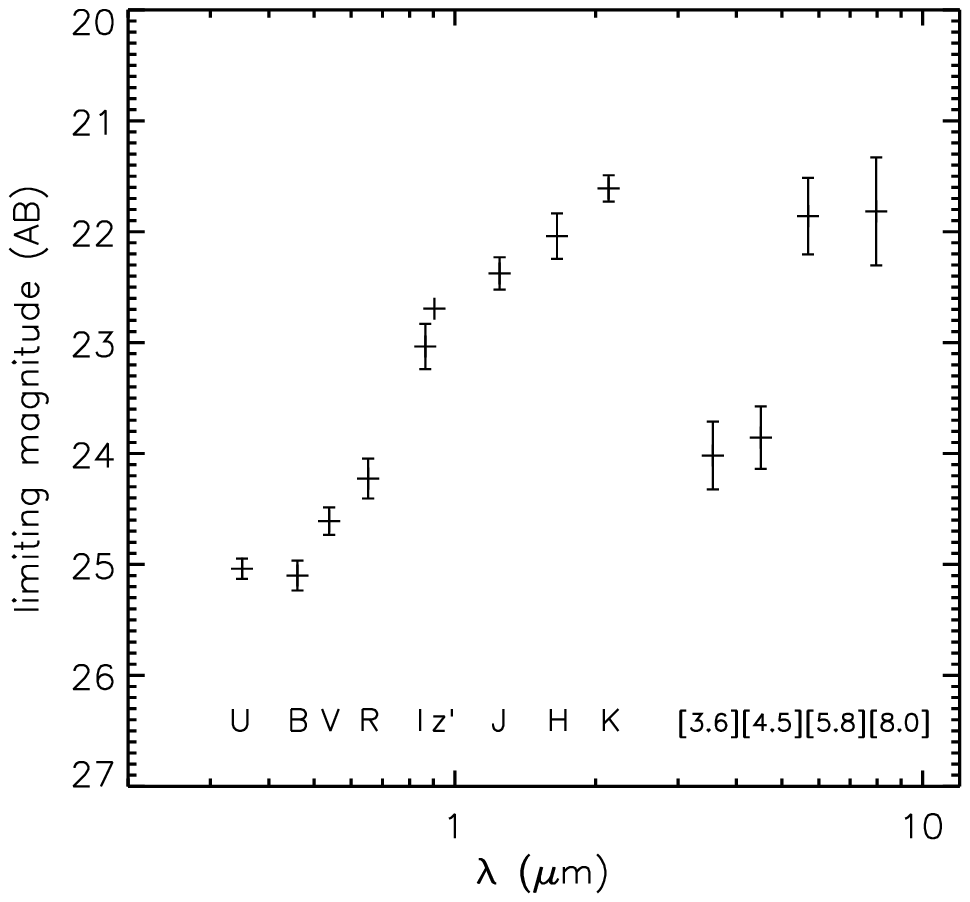}
   \caption[]{Limiting magnitude vs. bandpass wavelength in the SIMPLE
     catalog. The limiting depths are 5-$\sigma$ total magnitudes of
     point sources measured in apertures with a 2\arcsec.0
     radius. Since the exposure time varies for each band, there is
     scatter around each limiting magnitude. The error bars denote the 
     standard deviation of this scatter. Since we do not have an
     exposure map for the $z'$-band data, there is no error bar at the
     limiting magnitude of that band (see Taylor et al. 2009b). The
     IRAC magnitude limits have been determined excluding the GOODS
     data.
     \label{mag_lim}}
 \end{figure}
}

\def\figj{
  \begin{figure*}
    \includegraphics[width=\textwidth]{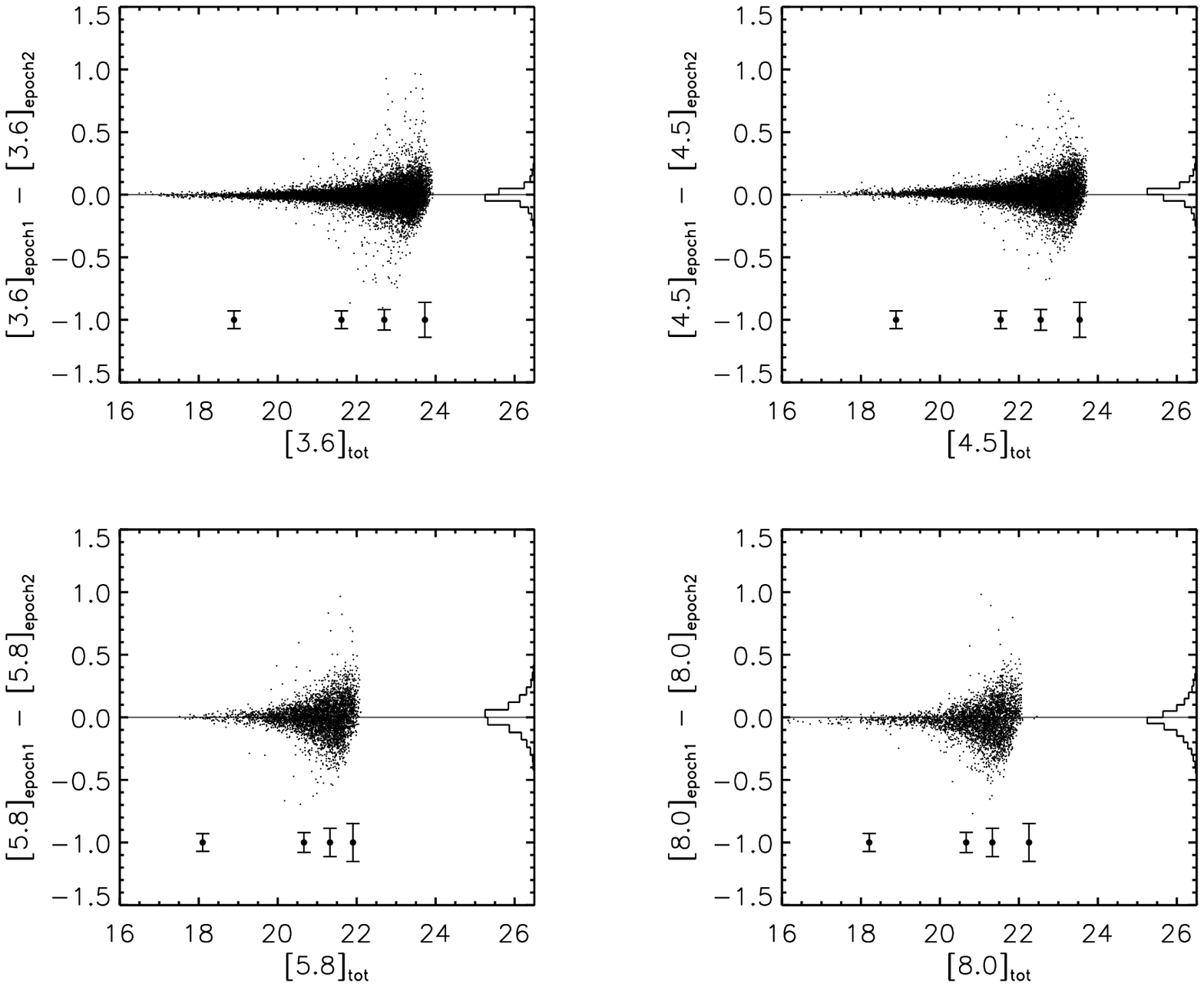}
    \caption[epochs]{Comparison between IRAC magnitudes of the first
      and second epoch of observations. The panels show the difference
      between the measured magnitudes of the four IRAC bands. At the
      right side of each panel, a histogram shows the distribution of
      the difference. The error bars are the mean errors in bins of
      equal number of sources, offset by -1 with respect to the
      measurements.
      \label{ep1ep2}}
  \end{figure*}
  }

\def\figk{
  \begin{figure}
    \includegraphics[width=0.5\textwidth]{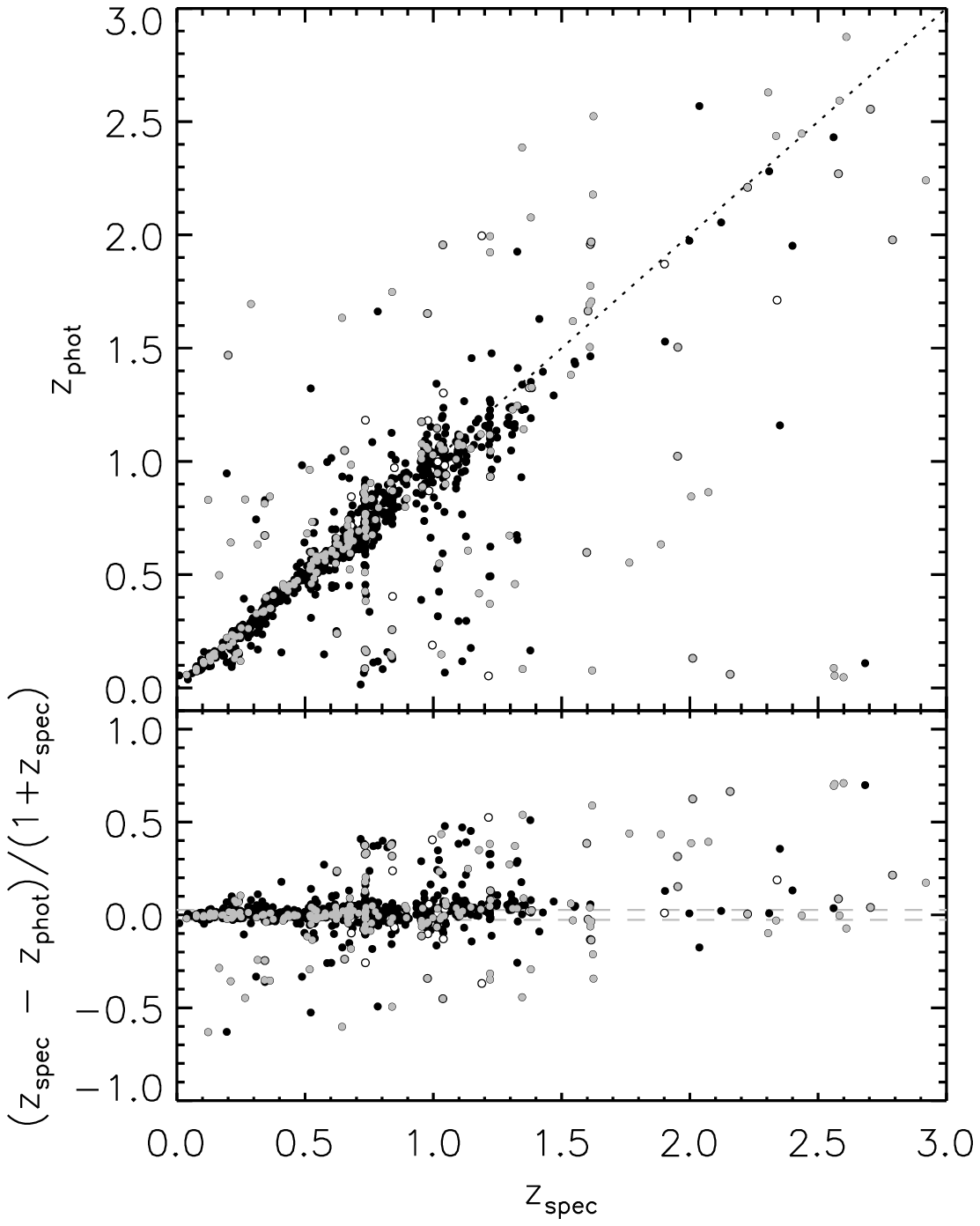}
    \caption[dz]{Photometric and spectroscopic redshifts in the
      E-CDFS. {\it Upper panel} - Direct comparison between
      photometric and spectroscopic redshifts for 1,226 IRAC detected
      sources with reliable $z_{spec}$ identification and coverage in
      all wavelength bands. The dotted line represents a one-to-one
      relationship. {\it Lower panel} - Residuals $dz = z_{spec}-
      z_{phot}/ ( 1 + z_{spec})$ as a function of spectroscopic
      redshift. The $\sigma_{NMAD}$ is 0.025, indicated by the dashed lines. Open
      circles denote AGN candidates, identified by their X-ray flux.
      \label{photz}}
  \end{figure}
}

\def\figl{
  \begin{figure*}
    \includegraphics[width=\textwidth]{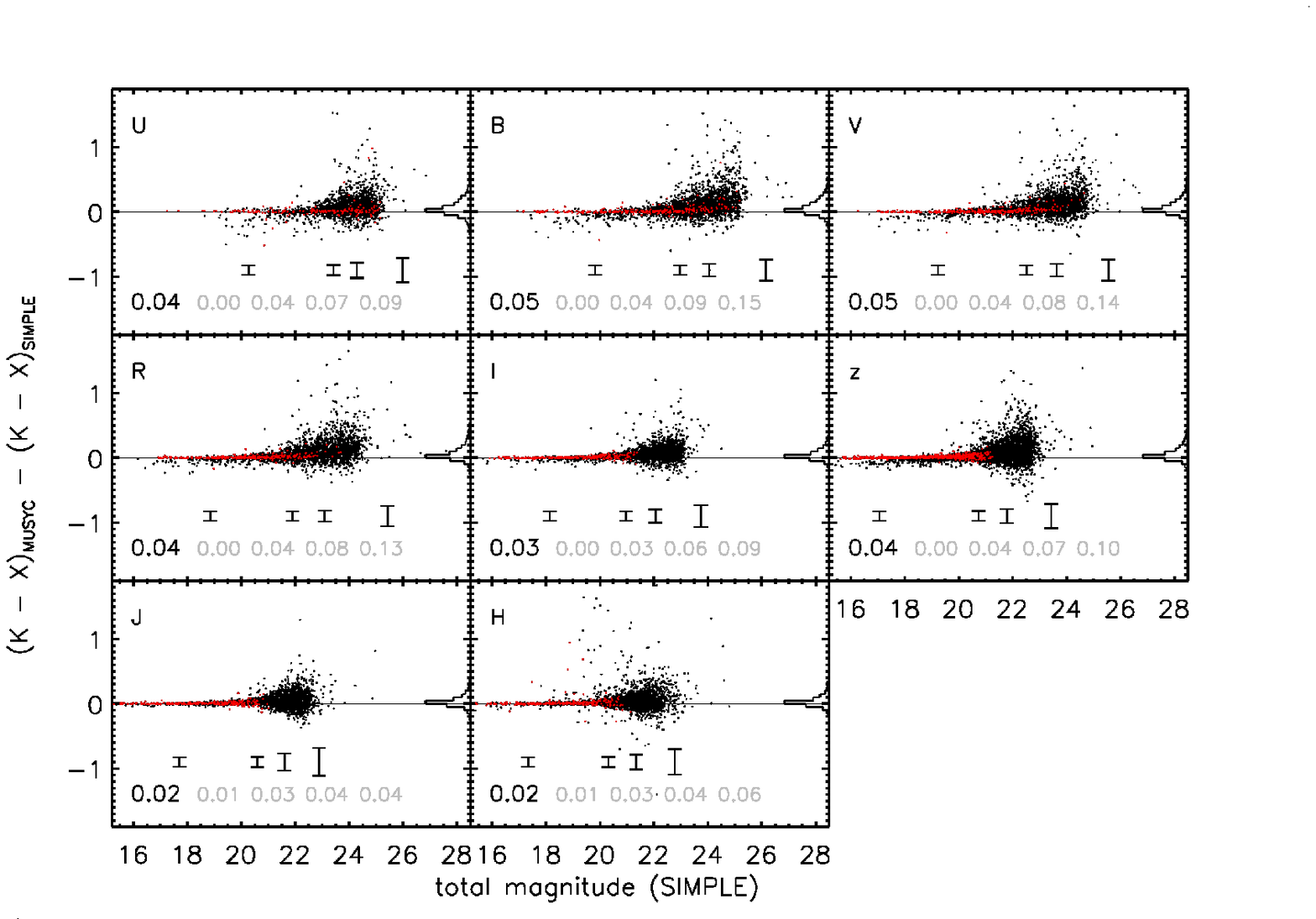}
    \caption[musyc]{Direct comparison between MUSYC and SIMPLE colors
      in the overlapping bands ($U-K$) for sources with S/N $ > 10$ in
      SIMPLE $K$-band. At the right side of each panel, a histogram
      shows the distribution of the offsets. Stars are shown in
      red. The median offset is indicated at the lower left corner of
      each panel. For each band only the SIMPLE sources with S/N $ > 5$
      are included. The error bars indicate the formal errors expected
      from the SIMPLE and MUSYC photometric errors. They are mean
      values in bins of equal number of sources and are offset by -1
      with respect to the measurements.
      \label{simple_musyc}}
  \end{figure*}
}

\def\figm{
  \begin{figure*}[p] 
    \includegraphics[width=\textwidth]{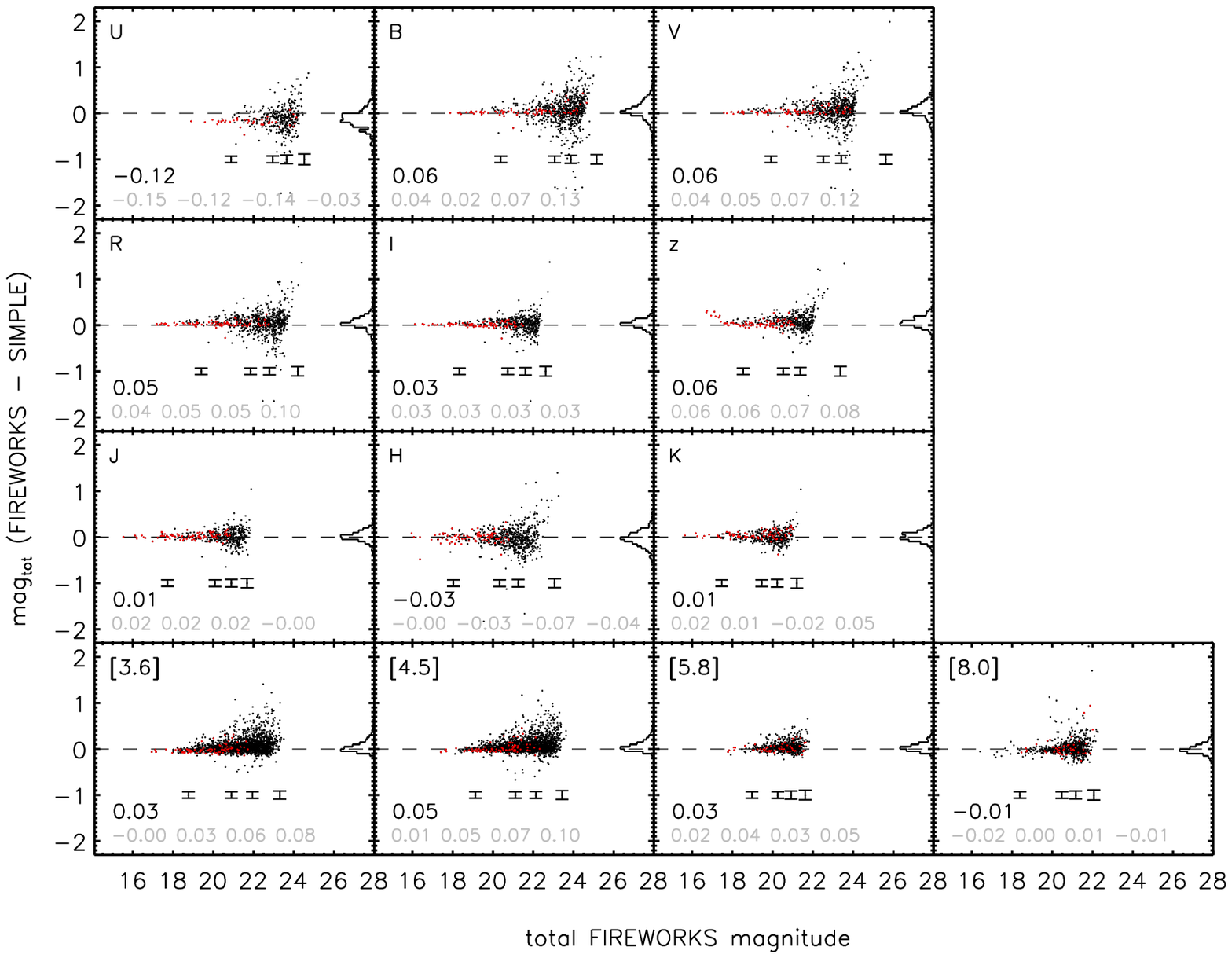}
    \caption[fw]{Direct comparison of total magnitudes for sources
      with S/N $ > 10$ at 4.5 \um\, for the $U-K$ + IRAC bands of the
      FIREWORKS catalog and our SIMPLE catalog. At the right side of
      each panel, a histogram shows the distribution of the
      offsets. The median offset is indicated at the lower left corner
      of each panel. For each band only the SIMPLE sources with S/N $ >
      10$ are included.  The error bars indicate the formal errors
      expected from the SIMPLE and FIREWORKS photometric errors. They
      are mean values in bins of equal number of sources and are
      offset by -1.5 with respect to the measurements. All blended
      FIREWORKS sources have been removed from this figure.
      \label{simple_fw}}
  \end{figure*}
}
\def\figii{
 \begin{figure}
   \includegraphics[width=0.5\textwidth]{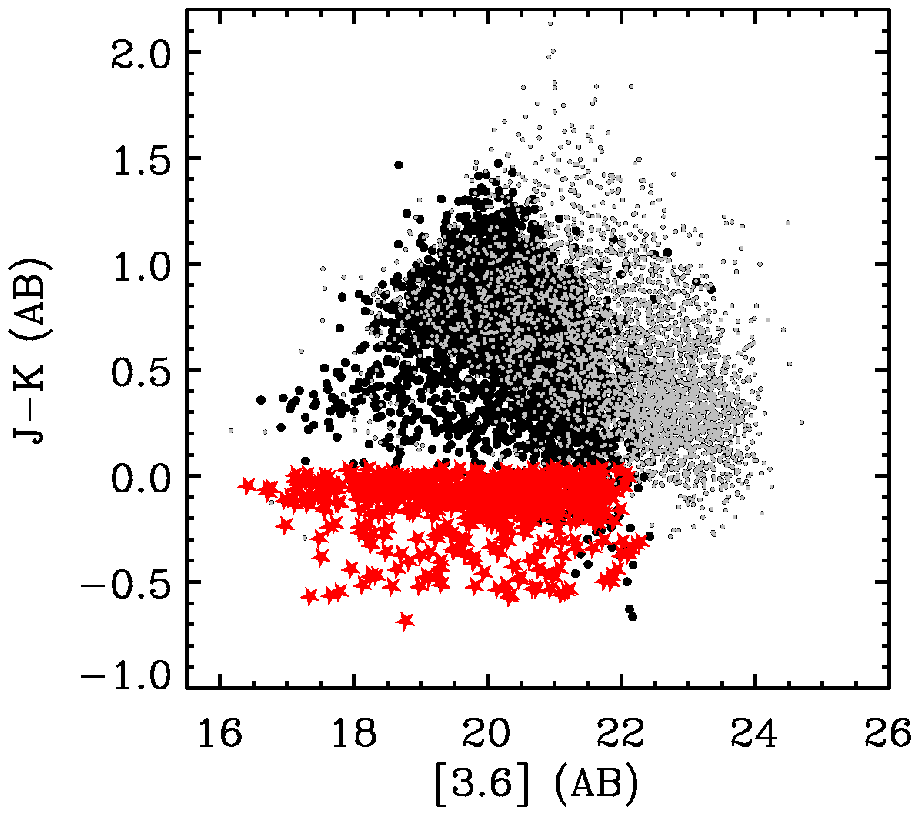}
   \caption[]{$J$-$K$ vs. [3.6] color-magnitude
      diagram for IRAC-selected sources in the E-CDFS. Stars ({\it red
        stars}) are identified by their $J$-$K$ color (see Section
      \ref{stars}). Overplotted in gray are the values from the
      FIREWORKS catalog, which reaches greater depth, but contains
      fewer sources out to a magnitude of 21.5. All blended FIREWORKS
      sources have been removed from this figure.
     \label{col_mag}}
 \end{figure}
}

\def\figo{
  \begin{figure*}[t!] 
    \includegraphics[width=\textwidth]{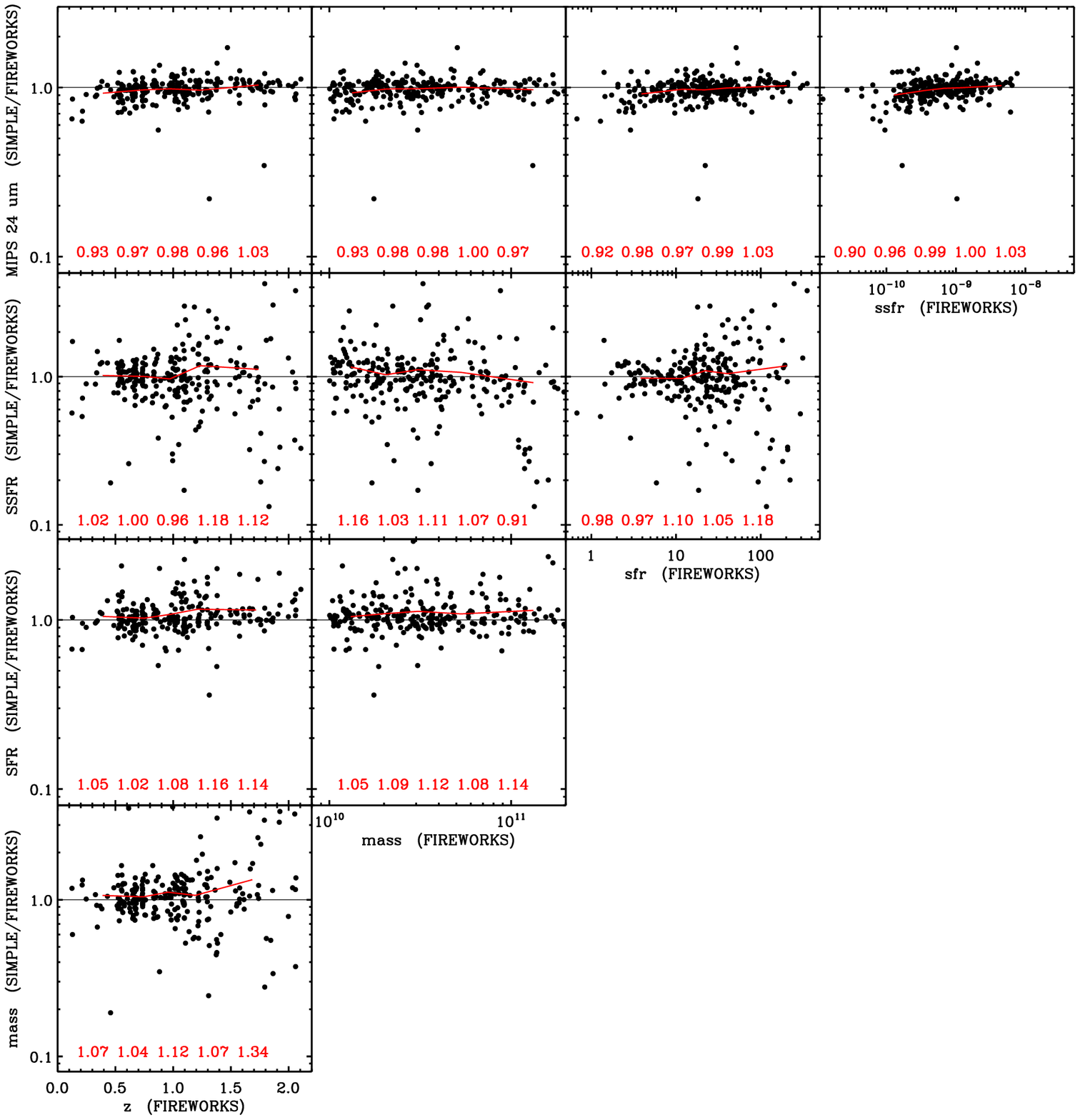}
    \caption[fw]{Comparison between various observed or derived
      quantities in FIREWORKS and SIMPLE. The red line indicates a
      binned mean of the difference in each quantity. The mean values
      are derived for five intervals of equal number of sources and
      are shown in red at the bottom of each panel. All blended
      FIREWORKS sources have been removed from this figure.
      \label{fw_other}}
  \end{figure*}
}

\def\figp{
  \begin{figure}
    \includegraphics[width=0.5\textwidth]{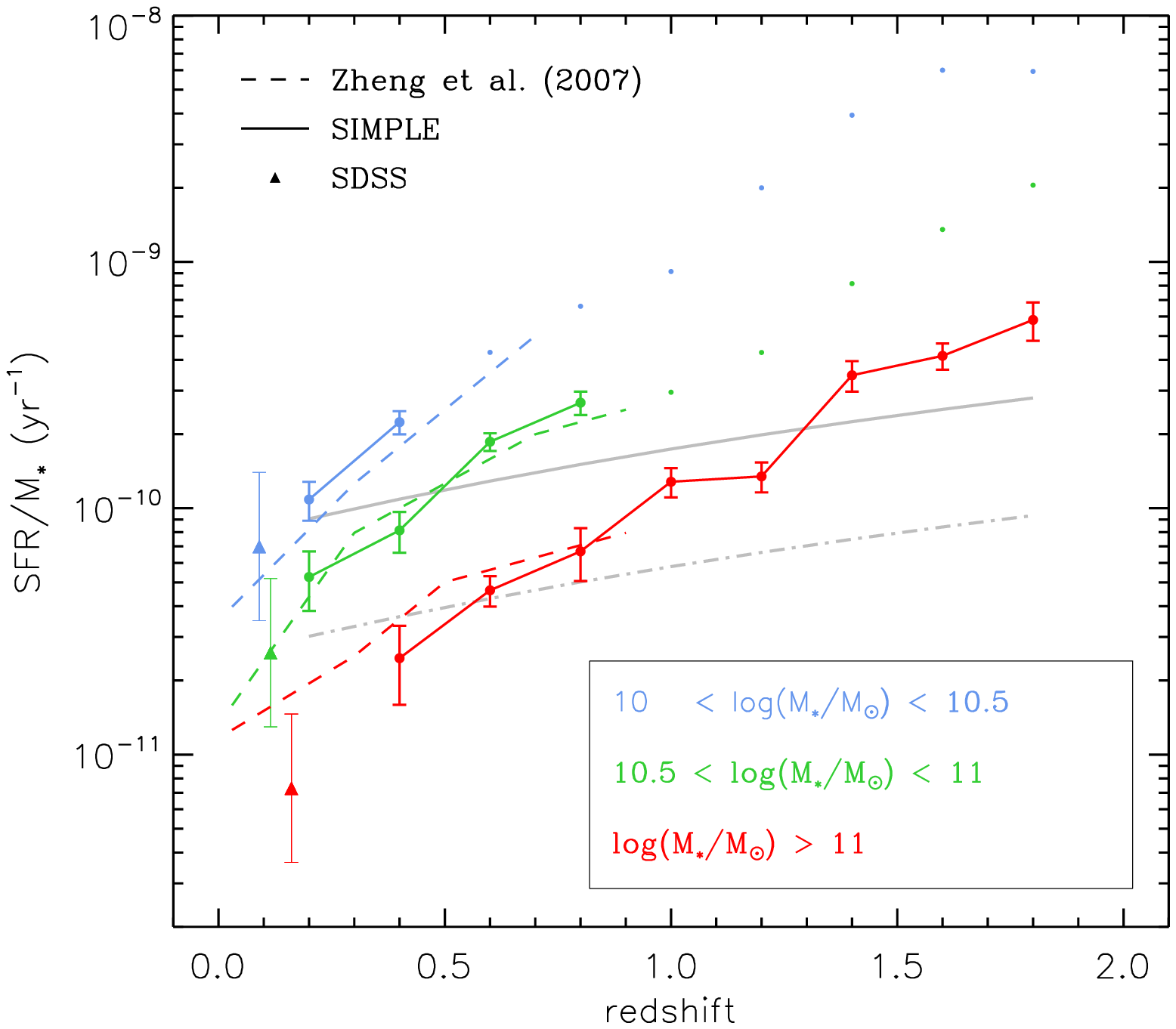}
    \caption[colmagfd]{Specific star formation rate versus
      redshift. This figure is an updated version of Figure 3 of Damen
      et al. (2009a), based on improved total fluxes. Filled circles
      are SIMPLE results, dots show where we become incomplete with
      respect to mass. Triangles denote SDSS data. The error bars
      represent bootstrap errors for SIMPLE and a systematic error of
      0.3 dex for the SDSS data. The dashed colored lines represent
      the results from Zheng et al. (2007) in identical mass bins. The
      gray solid line is the inverse of the Hubble time ($1/t_{H}$ in
      $yr^{-1}$). Sources above this line are in starburst mode. The
      time it would take to produce the current stellar mass at the
      current SFR is smaller than a Hubble time. Star formation is
      quenched in galaxies under the gray dashed line ($1/(3\times
      t_{H})$); the bulk of their stars has already been formed. The
      sSFR increases with $z$ at a rate that appears independent of
      mass and sSFRs of more massive galaxies are typically lower than
      those of less massive galaxies over the whole redshift range.
      \label{ssfrz}} 
  \end{figure}
}

\def\figq{
  \begin{figure}[htbp]
    \includegraphics[width=0.4\textwidth]{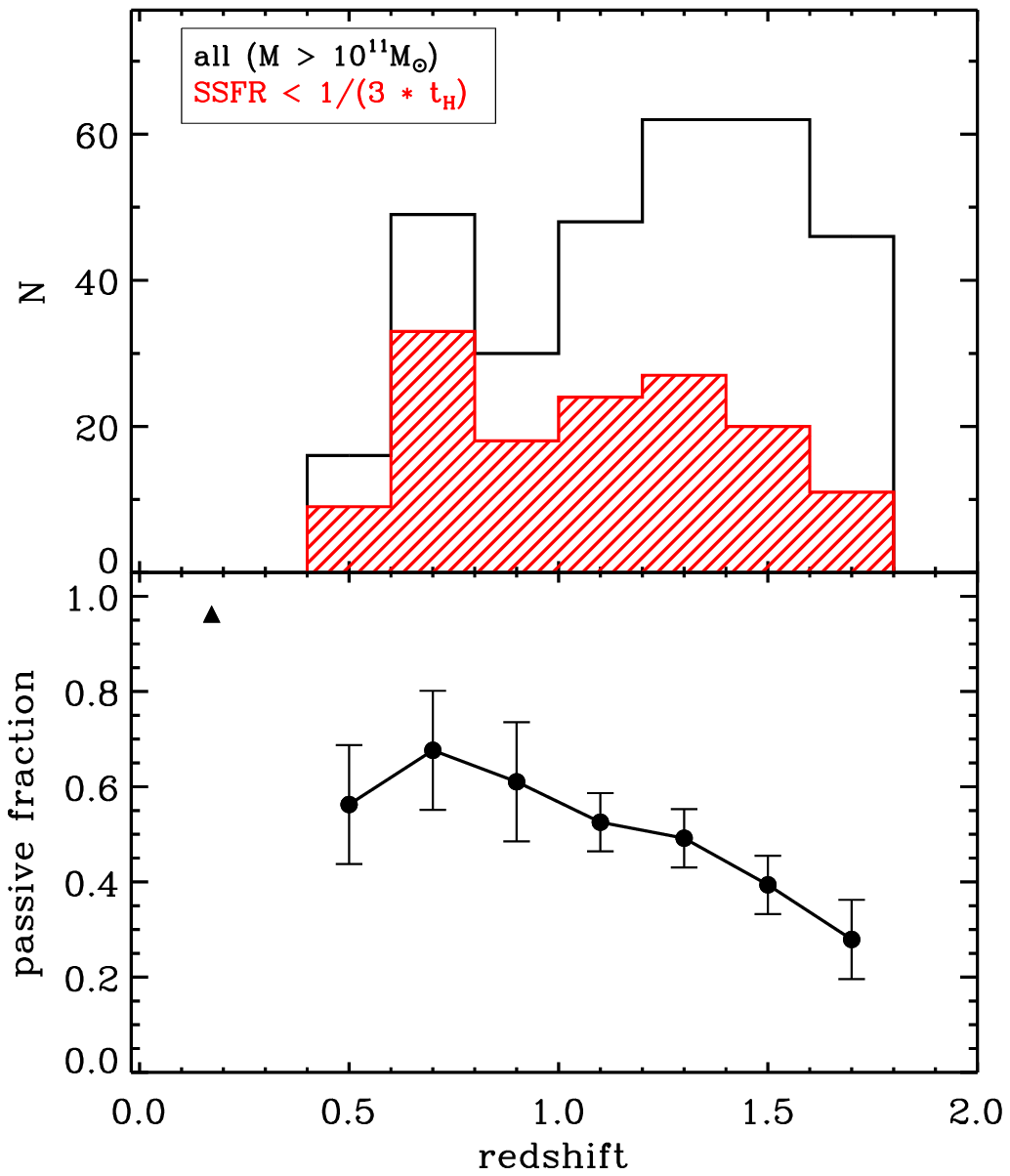}
    \caption[colmagfd]{The redshift evolution of the fraction of
      quiescent galaxies with $M_* > 10^{11} \Msol$. This figure is an
      updated version of Figure 6 of Damen et al. (2009a), based on
      improved total fluxes. The numbers are consistent with each
      other within 1-$\sigma$. Quiescent galaxies are defined as sources
      with sSFR$< 1/(3 \times t_H ) $\, yr$^{-1}$, where $t_H$ is the
      age of the universe at the source's redshift. The upper panel
      shows the distribution of all redshift in this mass range,
      overplotted is the number of galaxies whose star formation is
      quenched. The lower panel shows the fraction of galaxies in
      quiescent mode. The error bars represent bootstrap errors. SDSS
      data have been used to determine a local values ({\it
        triangle}). 
      \label{frac}}
  \end{figure}
}

\def\figaa{
  \begin{figure}
    \includegraphics[width=\textwidth]{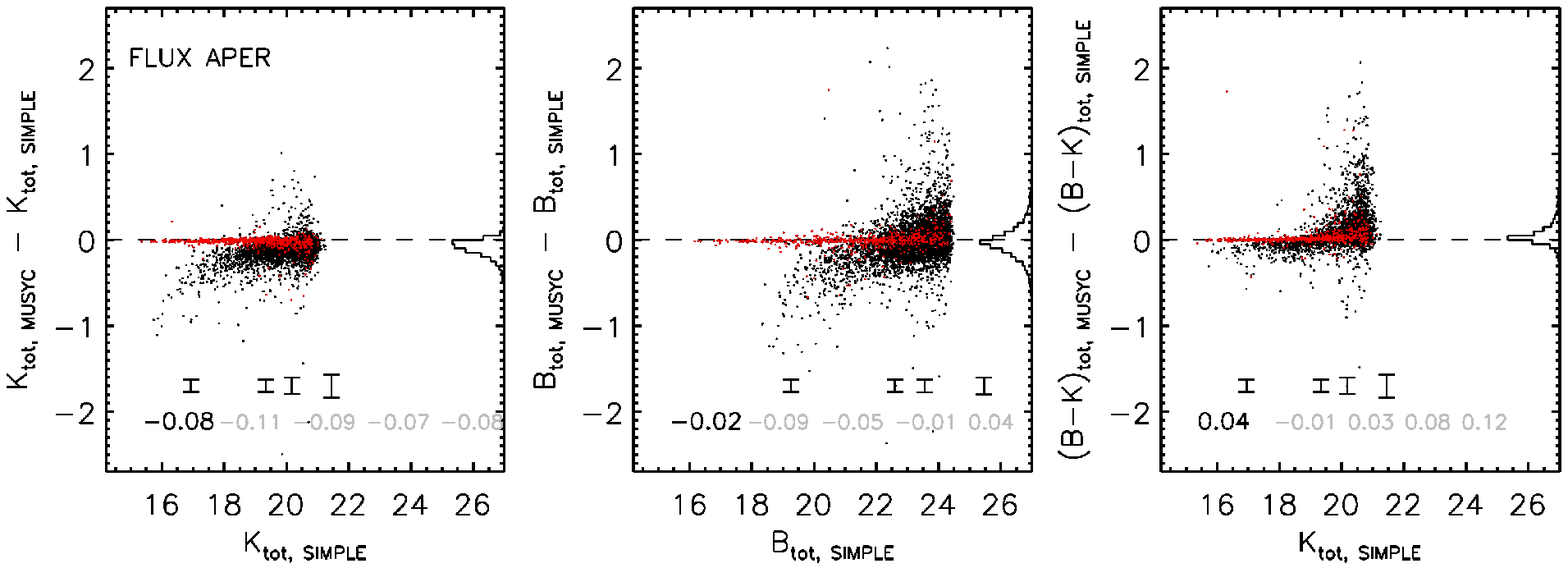}
    \includegraphics[width=\textwidth]{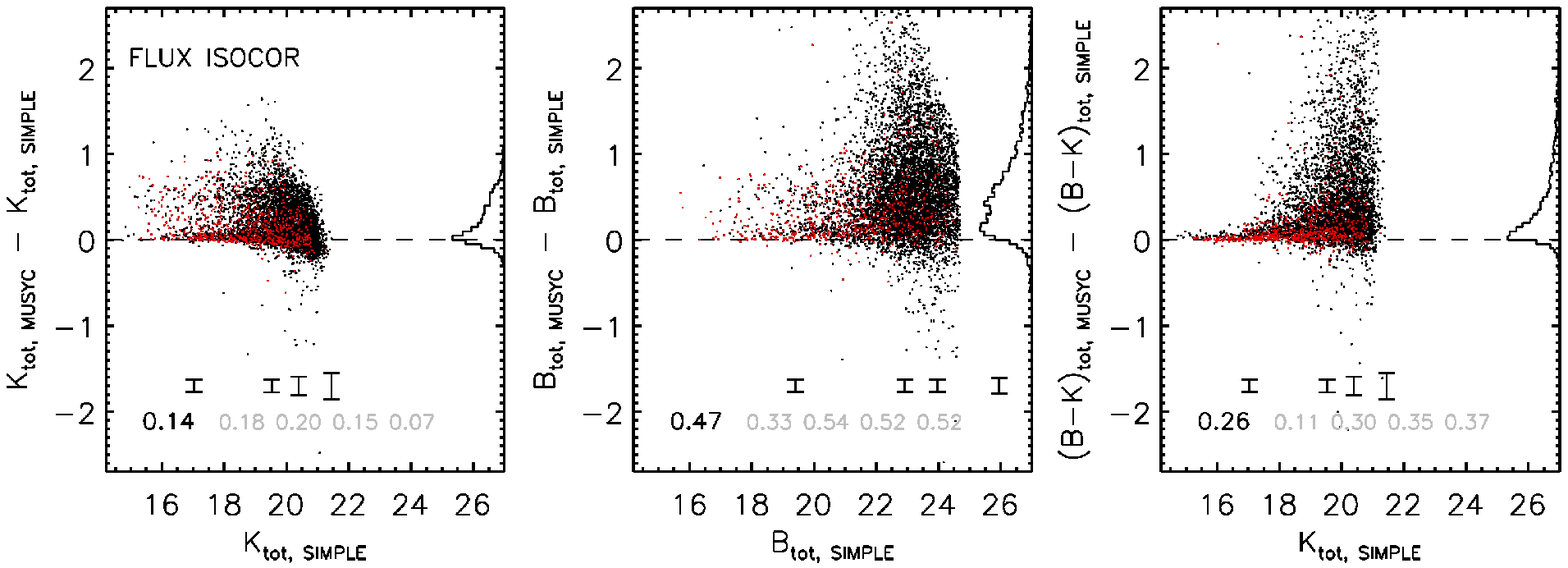}
    \includegraphics[width=\textwidth]{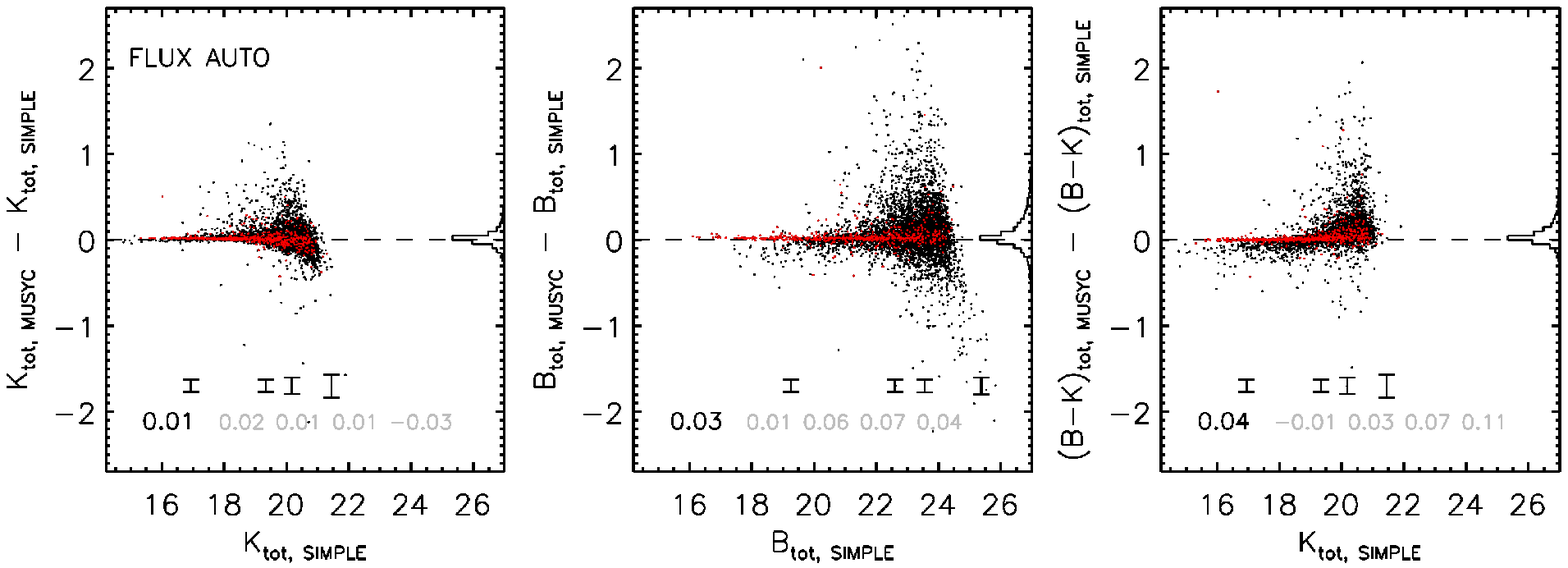}
    \caption[]{Comparison between MUSYC and SIMPLE magnitudes in the
      $B$- and $K$-bands for different apertures. The apertures used
      are (from top to bottom row) fixed apertures of 4\arcsec\, (FLUX
      APER), corrected isophotal apertures (FLUX ISOCOR), and flexible
      elliptical apertures (FLUX AUTO). Stars are shown in
      red. The median offset is indicated at the lower left corner of
      each panel. For each band only the SIMPLE sources with S/N $ > 5$
      are included. The error bars indicate the formal errors expected
      from the SIMPLE and MUSYC photometric errors. They are mean
      values in bins of equal number of sources and are offset by -1
      with respect to the measurements. In the construction of this
      figure no distinction has been made between blended and
      non-blended sources.
      \label{fluxapers}}
  \end{figure}
}

\def\figab{
    \begin{figure}
    \includegraphics[width=0.4\textwidth]{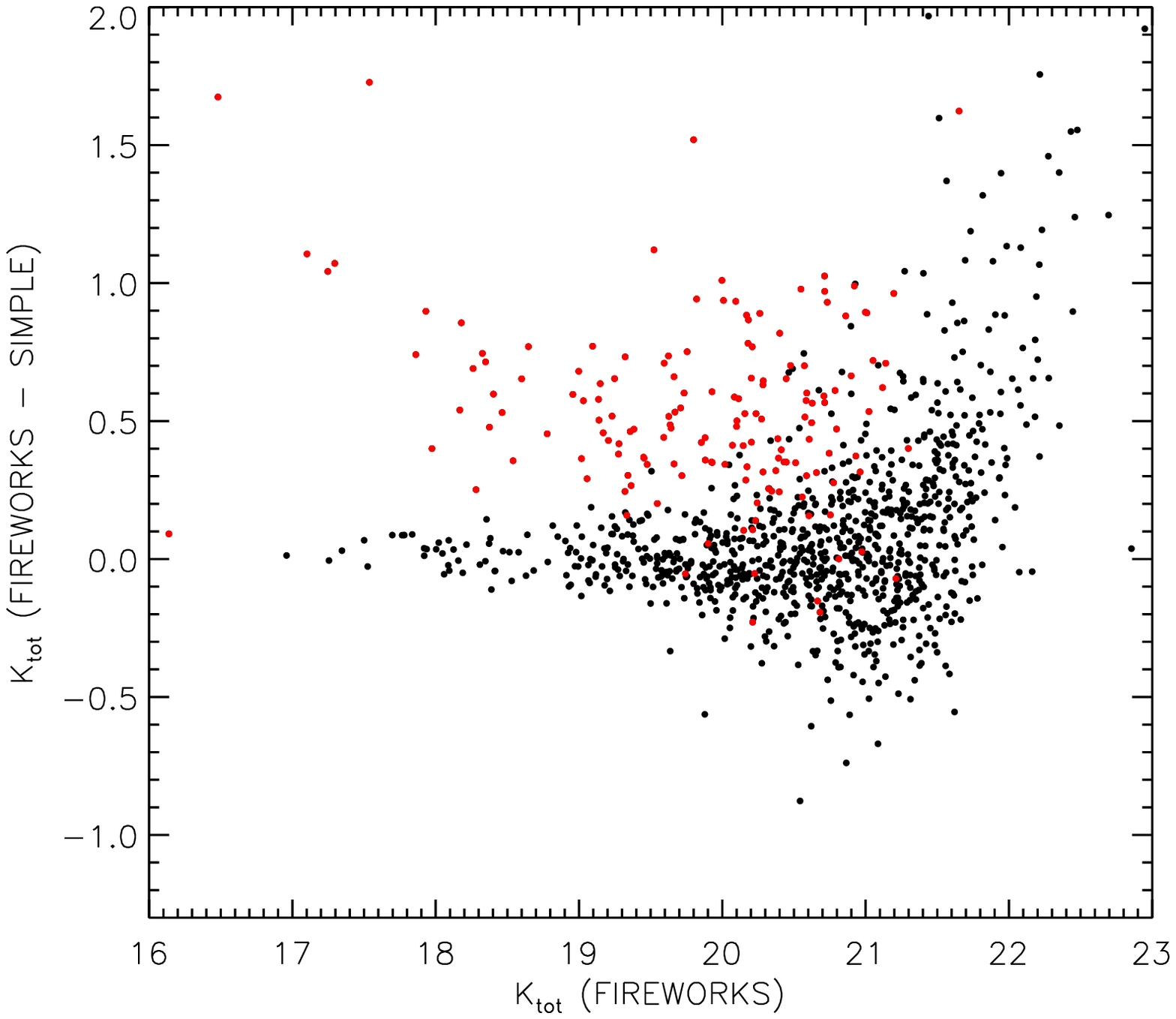}
    \caption[]{Comparison between FIREWORKS and SIMPLE for $K$-band
      total magnitude. Sources that are blended in the FIREWORKS
      catalog are shown in red. We removed these sources from all
      analysis (i.e., Figures \ref{simple_fw}-\ref{fw_other}). The
      sources that are flagged as blended by SExtractor take up
      $\gtrsim$60\% of the complete SIMPLE sample and even a higher
      fraction (98\%) of the sources shown above, which are relatively
      bright ((S/N)$_K >$ 5).
      \label{kblend}}
  \end{figure}
}

\def\figac{
  \begin{figure}
    \includegraphics[width=0.4\textwidth]{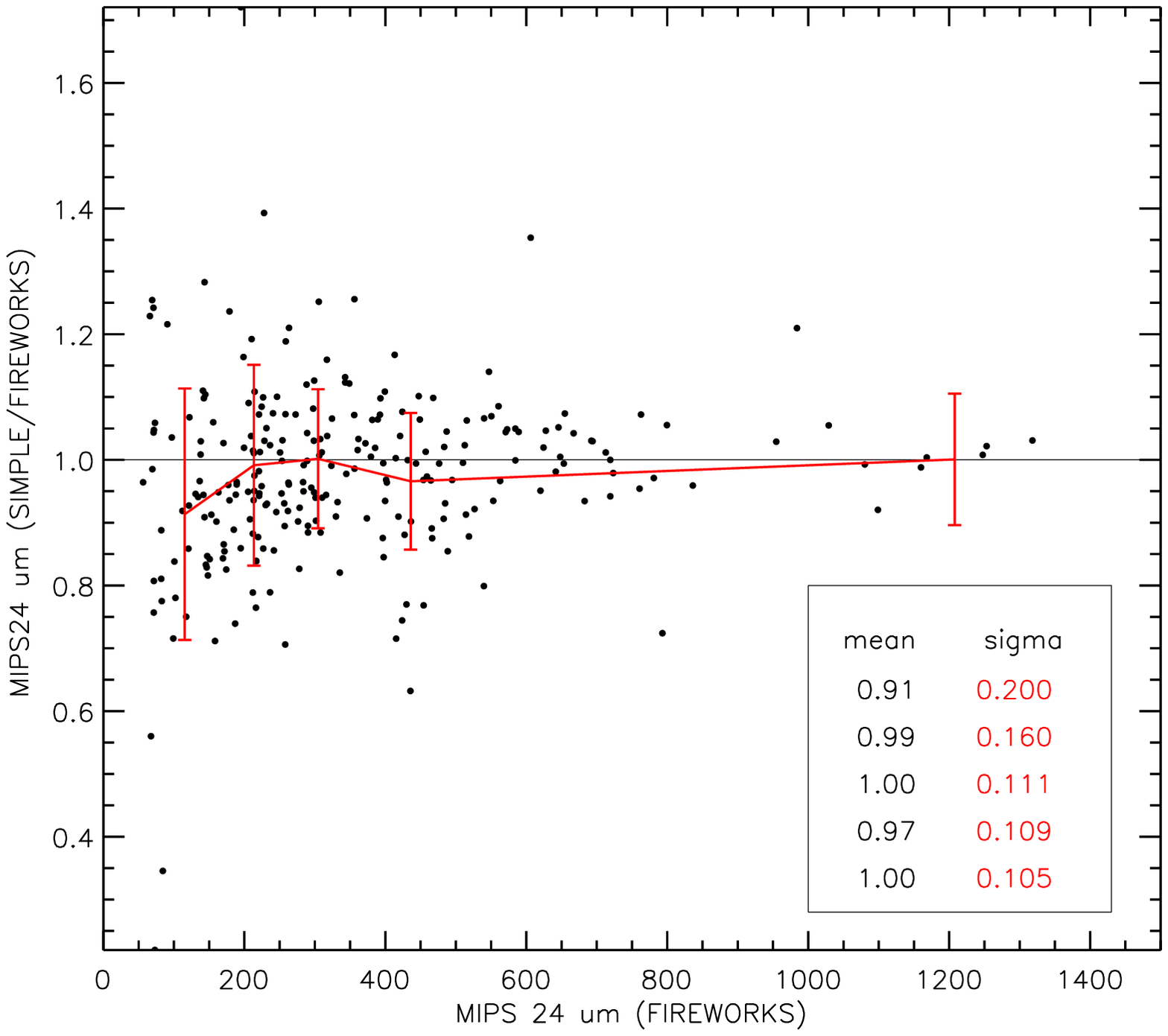}
    \caption[]{Comparison between FIREWORKS and SIMPLE for MIPS 24
      \um\, total magnitude. The mean values are indicated by the red
      line and printed in the lower right corner, together with the
      standard deviation in each bin. The formal errors obtained from
      our deblending routine are smaller than the observed standard
      deviation. It is clear, however, that the MIPS 24 \um\, fluxes
      are consistent with each other within 10\%-20\%. 
      \label{mips_scatter}}
  \end{figure}
  }

\def\figad{
  \begin{figure}
    \includegraphics[width=\textwidth]{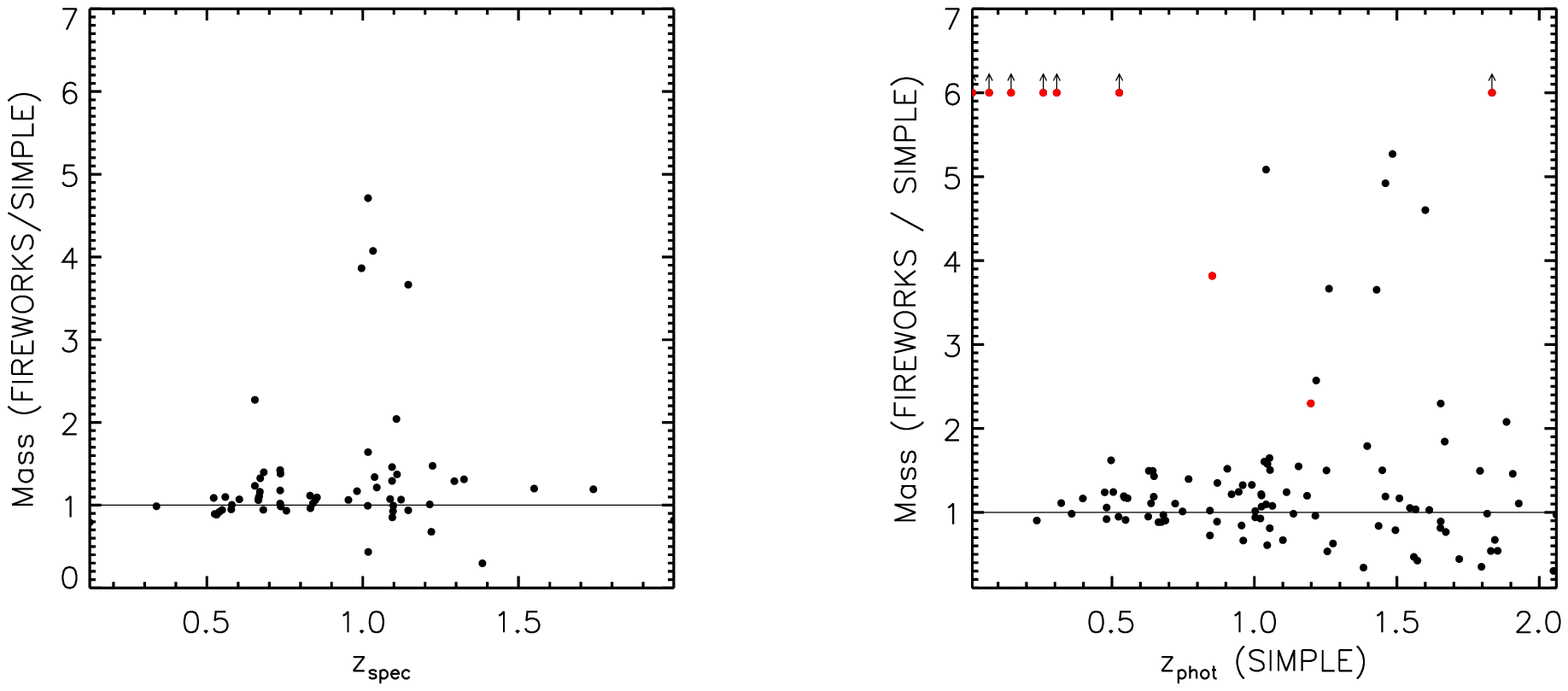}
    \caption[]{Comparison between the masses of FIREWORKS and SIMPLE
      versus (a) spectroscopic and (b) photometric redshifts. The red
      points in the right panel represent sources with photometric
      redshifts that differ more than 0.5 between FIREWORKS and
      SIMPLE. Despite these outliers, it is clear that the observed
      scatter is not caused by photometric redshift errors.
      \label{zphot_mass}}
  \end{figure}
}

\begin{document}

\title{The SIMPLE survey: observations, reduction, and catalog}

\author{M. Damen\altaffilmark{1}, I. Labb\'e\altaffilmark{2},
  P.~G. van Dokkum\altaffilmark{3}, M. Franx\altaffilmark{1},
  E. N. Taylor\altaffilmark{4}, W.~N. Brandt\altaffilmark{5},
  M. Dickinson\altaffilmark{6}, E. Gawiser\altaffilmark{7},
  G.~D. Illingworth\altaffilmark{8}, M. Kriek\altaffilmark{9},
  D. Marchesini\altaffilmark{10}, A. Muzzin\altaffilmark{3},
  C. Papovich\altaffilmark{11}, H.-W. Rix\altaffilmark{12}}

\email{damen@strw.leidenuniv.nl}

\altaffiltext{1}{Leiden Observatory, Leiden University, PO Box 9513,
  2300 RA Leiden, The Netherlands}
\altaffiltext{2}{Carnegie Observatories, 813 Santa Barbara Street,
  Pasadena, CA 91101, USA; Hubble Fellow}
\altaffiltext{3}{Department of Astronomy, Yale University, New Haven,
  CT, 06520, USA}
\altaffiltext{4}{School of Physics, The University of Melbourne,
  Parkville, 3010, Australia}
\altaffiltext{5}{Department of Astronomy and Astrophysics, The
  Pennsylvania State University, University Park, PA 16802, USA}
\altaffiltext{6}{NOAO, 950 N. Cherry Avenue, Tucson, AZ 85719, USA}
\altaffiltext{7}{Department of Physics and Astronomy, Rutgers
  University, Piscataway, NJ 08854, USA}
\altaffiltext{8}{UCO/Lick Observatory, University of California, Santa
  Cruz, CA 95064, USA}
\altaffiltext{9}{Harvard-Smithsonian Center for Astrophysics, 60 Garden 
Street, Cambridge, MA 02138}
\altaffiltext{10}{Department of Physics and Astronomy, Tufts
  University, Medford, MA 02155, USA}
\altaffiltext{11}{George P. and Cynthia Woods Mitchell Institute for
  Fundamental Physics and Astronomy, Department of Physics and
  Astronomy, Texas A\&M University, 4242 TAMU, College Station, TX
  77843, USA}
\altaffiltext{12}{Max-Planck-Institut f\"ur Astronomie, K\"onigstuhl
  17, D-69117 Heidelberg, Germany}

\begin{abstract}
We present the {\it Spitzer} IRAC/MUSYC Public Legacy Survey in the
Extended CDF-South (SIMPLE), which consists of deep IRAC observations
covering the $\sim$1,600 arcmin$^2$ area surrounding GOODS-S. The
limiting magnitudes of the SIMPLE IRAC mosaics typically are 23.8,
23.6, 21.9, and 21.7, at 3.6 \um, 4.5 \um, 5.8 \um, and 8.0 \um,
respectively (5-$\sigma$ total point source magnitudes in AB). The
SIMPLE IRAC images are combined with the 10\arcmin \, $\times$ 15\arcmin
\,GOODS IRAC mosaics in the center. We give detailed descriptions of
the observations, data reduction, and properties of the final images,
as well as the detection and photometry methods used to build a
catalog. Using published optical and near-infrared data from the
Multiwavelength Survey by Yale-Chile (MUSYC), we construct an
IRAC-selected catalog, containing photometry in $UBVRIz'JHK$, [3.6
\um], [4.5 \um], [5.8 \um], and [8.0 \um]. The catalog contains 43,782
sources with S/N $> 5$ at 3.6 \um, 19,993 of which have 13-band
photometry. We compare this catalog to the publicly available MUSYC
and FIREWORKS catalogs and discuss the differences. Using a high
signal-to-noise sub-sample of 3,391 sources with $ (\mone + \mtwo)/2 <
21.2$, we investigate the star formation rate history of massive galaxies out to $z
\sim 1.8$. We find that at $z \sim 1.8$ at least 30\%$\pm7$\% of the most
massive galaxies (\Mstar\, $> 10^{11}$\Msol) are passively evolving,
in agreement with earlier results from surveys covering less area. 
\end{abstract}


\section{Introduction}
Our understanding of galaxy formation and evolution has dramatically
increased through the rise of large and deep galaxy surveys that have
opened up the high-redshift universe for research. The best studied
high-redshift galaxies are arguably the Lyman break galaxies (LBGs)
that can be identified by their rest-frame UV colors (Steidel et
al. 1996; 1999). Although much has been learned from studying their
properties, LBGs are not representative for all high-redshift galaxy
populations. \\
Since they are based on selection in the rest-frame UV, optical
surveys of high-redshift galaxies are heavily affected by dust
obscuration and are not sensitive to old stellar populations. The
rest-frame optical is less influenced by the contribution from young
stars and dust and provides a more reliable means of tracing the bulk
of the stellar mass at high redshift. For instance, near-infrared
observations have uncovered a significant population of massive, red
galaxies, particularly at high redshift (Elston, Rieke \& Rieke 1988,
Spinrad et al. 1997, Barger et al. 1999, Daddi et al. 2000, Franx et
al. 2003, Labb\'e et al. 2003, Cimatti et al. 2004, van Dokkum et
al. 2006).\\
With the arrival of the {\it Spitzer Space Telescope} and its Infrared
Array Camera (IRAC; Fazio et al. 2004), constructing large surveys to
study high-redshift galaxies has become even more attainable, since
the IRAC wavelengths provide coverage of the rest-frame optical bands
out to higher redshifts. Using deep IRAC observations at 4.5 \um\, it
is possible to trace the rest-frame $I$-band out to a redshift $z
\sim$ 4.\\

 \begin{deluxetable*}{lcccll}
     \tablecolumns{6}
     \tablewidth{0pc}
     \tablecaption{N-$\sigma$ limiting depths (total AB
       magnitude)\label{spitzer.tab}
     }
     \tablehead{
       \colhead{} & \colhead{}
     }
     \startdata
     Program   & area              & channel  & depth (AB mag) & S/N &
     integration time\\
     GOODS-S   &  138 arcmin$^2$   & 3.6 \um  & 26.15  & 3 & 23 hr\\
               &                   & 4.5 \um  & 25.66  &   &    \\
               &                   & 5.8 \um  & 23.79  &   &    \\
               &                   & 8.0 \um  & 23.70  &   &    \\
     SIMPLE    &  1,600 arcmin$^2$ & 3.6 \um  & 23.86  & 5 & 0.9-2.5
     hr\\
               &                   & 4.5 \um  & 23.69  &   &    \\
               &                   & 5.8 \um  & 21.95  &   &    \\
               &                   & 8.0 \um  & 21.84  &   &    \\
     S-COSMOS  &  2 deg$^2$        & 3.6 \um  & 24.0   & 5 & 1200 s\\
               &                   & 4.5 \um  & 23.3   &   &    \\
               &                   & 5.8 \um  & 21.3   &   &    \\
               &                   & 8.0 \um  & 21.0   &   &    \\
     SWIRE     &  60 deg$^2$       & 3.6 \um  & 21.4   & 10& 120-480
     s\\
               &                   & 4.5 \um  & 21.4   & 5 &    \\
               &                   & 5.8 \um  & 19.8   &   &    \\
               &                   & 8.0 \um  & 19.9   &   &    \\
     \enddata
   \end{deluxetable*}

The massive, red galaxies found at high redshift are important
test-beds for models of galaxy formation and evolution. To be able to
place constraints on the models we need a clear picture of the
evolution and star formation history of these massive galaxies. This
requires large, statistically powerful samples, or in other words,
surveys that extend over a great area and depth.\\
It is also critical to do these observations in areas that already
have been observed at many wavelengths and ideally in areas that are
accessible to future telescopes such as ALMA. The 30\arcmin \, $\times$
30\arcmin\, Extended Chandra Deep Field South (E-CDFS) is perfect in
this sense as it is one of the most extensively observed fields
available. There is a large set of ground-based data providing
$UBVRIz'JHK$ imaging (MUSYC (Gawiser et al. 2006, Quadri et al. 2007,
Taylor et al. 2009b), COMBO-17 (Wolf et al. 2004), LCIRS, (McCarthy et
al. 2001)), radio coverage (Miller et al. 2008), and spectroscopy
(e.g., GOODS (VIMOS: Popesso et al. 2009, FORS2: Szokoly et al. 2004,
Vanzella et al. 2008), MUSYC (Treister et al. 2009a), K20 (Cimatti et
al. 2002), VVDS (le F\`evre et al. 2004)). The area has been targeted
intensely from space too. There is HST ACS imaging from GEMS (Rix et
al. 2004), observations from CHANDRA (Lehmer et al. 2005, Luo et
al. 2008), XMM (PI: J. Bergeron), GALEX (Martin et al. 2005), and
ultra deep multiwavelength coverage from the Great Observatories
Origins Deep Survey (GOODS, Dickinson et al. 2003) in the
central 10\arcmin \, $\times$ 15\arcmin. The rich multiwavelength
coverage includes also deep 24, 70, and 160 \um\, observations from the Far-Infrared
Deep Extragalactic Legacy Survey (FIDEL).\\
In this context, we initiated {\it Spitzer}'s IRAC + MUSYC Public
Legacy of the E-CDFS (SIMPLE), which aims to provide deep, public IRAC
imaging of a 1,600 arcmin$^2$ area on the sky. In this paper, we
present the full IRAC data set, with an IRAC-selected multicolor
catalog of sources with 13-band optical-to-infrared photometry
(covering 0.3-8.0 \um). The optical to near-infrared (NIR) data come
from the Multiwavelength Survey by Yale-Chile (MUSYC; Taylor et
al. 2009), which are publicly
available\footnote{http://www.astro.yale.edu/MUSYC}. We also included
the 24 \um\, data from FIDEL, which reaches a depth of $\sim 40 \uJy$.\\
In addition to the study of massive galaxies, the SIMPLE survey can be
used to analyze properties of active galactic nuclei (AGNs). Luminous optically unobscured AGN can
be selected based on their IRAC colors (Lacy et al. 2005, Stern et
al. 2005). In the case of dust-obscured AGNs, the energy absorbed at
optical to X-ray wavelengths is later re-emitted in the mid-IR. AGNs
should therefore by very bright mid-IR sources (e.g., Daddi et al. 2007,
Alexander et al. 2008, Donley et al. 2008). The SIMPLE survey has
proved valuable in this context (Cardamone et al. 2008, Treister
et al. 2009a, 2009b) and the full photometric data set in the E-CDFS 
can provide strong constraints on the redshifts, masses, and stellar 
populations of the host galaxies. One example is the study by Luo et
al. (2010), who use the SIMPLE survey (among other data sets) to
improve photometric redshifts of X-ray AGN hosts. Their work also
shows the value of the SIMPLE survey regarding the identification of
X-ray selected AGNs. These sources can be very difficult to identify
at faint counterpart magnitudes. Luo et al. (2010) quantify a
counterpart recovery rate and found that the SIMPLE 3.6 \um\, data
score high in that respect (see their Table 1).
To conclude the list of SIMPLE applications, IRAC observations have
been useful in investigating the stellar populations of
Ly$\alpha$-emitting galaxies (Lai et al. 2008).\\ 
Here, we focus on the observations, data reduction processes, and the
construction of the catalog. This paper is structured as follows. In
Section \ref{obs}, we describe the observations with IRAC. Section
\ref{reduction} explains the reduction processes and the combined IRAC
mosaics. The ancillary data from the MUSYC and FIDEL surveys that we
use are described in Section \ref{ancil}. Source detection and
photometry are discussed in Section \ref{detphot}. In Section
\ref{redshifts}, we examine our photometric redshifts by comparing them
to a compilation of spectroscopic redshifts. The catalog parameters
are listed and explained in Section \ref{catalog} and Section
\ref{comp} describes the comparison of the SIMPLE catalog with two
other catalogs of the (E-)CDFS.\\ 
In a recent paper (Damen et al. 2009a), we used the SIMPLE catalog to
study the evolution of star formation in massive galaxies. Those
results were based on a preliminary version of the catalog and we
update the conclusions in Section \ref{ssfr}. Finally, Section
\ref{summary} provides a summary of the paper.\\
Throughout the paper we assume a $\Lambda$CDM cosmology with
$\Omega_{\rm m}=0.3$, $\Omega_{\rm  \Lambda}=0.7$, and $H_{\rm
  0}=70$~km s$^{-1}$ Mpc$^{-1}$. All magnitudes are given in the AB
photometric system. We denote magnitudes from the four \spitzer\ IRAC
channels as \mone, \mtwo, \mthree, and \mfour, respectively. Stellar
masses are determined assuming a Kroupa (2001) initial mass function
(IMF).\\

\section{Observations}\label{obs}
The SIMPLE IRAC Legacy survey consists of deep observations with IRAC (Fazio et al. 2004) covering the $\sim $
1600 arcmin$^2$ area centered on the GOODS-IRAC imaging (Dickinson et
al. 2003) of the CDFS (Giacconi et
al. 2002). The survey is complementary in area and depth to other
legacy programs, such as GOODS-IRAC (138 arcmin$^2$, 1380 minutes (Dickinson
et al. 2003)), S-COSMOS (2 deg$^2$, 20 minutes (Sanders et al. 2007)), the
Spitzer Wide-Area InfraRed Extragalactic Survey (SWIRE; 60 deg$^2$,
2-8 minutes (Lonsdale et al. 2003)), the Spitzer Ultra Deep Survey (SpUDS;
0.8 deg$^2$, $\sim$120 minutes (PI: J. Dunlop)), and the Spitzer
Deep, Wide-Field Survey (SDFWS; 10 deg$^2$, 6 minutes, including the
Irac Shallow Survey (ISS) (Ashby et al. 2009)) (see Table
\ref{spitzer.tab} for more details). The goal of 
the SIMPLE survey was to map a large area around the CDFS, with an
optimum overlap with existing surveys such as the GEMS project (Galaxy
Evolution from Morphology and SEDs), COMBO-17, and MUSYC. The area of
the CDFS appears as a hole in the center of the mosaic. The central
coordinates of the field are: $\alpha = 3^h32^m29.^s460, \delta =
-27\arcdeg 48\arcmin 18\arcsec.32$, J2000). Figure \ref{fov}
illustrates the field of the main surveys of the E-CDFS: GOODS (IRAC
and ACS), GEMS, COMBO-17, MUSYC, and SIMPLE.\\
\figa
The SIMPLE IRAC Legacy program was observed under program number GO
20708 (PI: van Dokkum). The complete set of observations consists of 36
pointings. On each pointing the mapping mode was used to observe a 2
$\times$ 3 rectangular grid, with each grid position receiving 30
minutes integration, for a total of 3 hr per pointing. The grand
total exposure time was $\sim$105 hr. The 2 $\times$ 3 map grids
partially overlap, leading to an average exposure time on the sky of
$\sim$1.5 hr. The observations were split in two epochs,
approximately 6 months apart.\\
\figb
The observations were split in two epochs, approximately 6 months
apart. The telescope orientation was rotated $\sim 170\degree$ between
the two epochs and this ensured that the area of the E-CDFS was fully
covered in all four IRAC bands. This is illustrated in
Fig. \ref{fov_ep}, which shows the exposure coverage of channel 1 (3.6
\um; left) and channel 2 (4.5 \um; right). Solid lines indicate the
outline of all observations from the first epoch, dashed lines those
of the second. IRAC observes in pairs: 3.6 and 5.8 \um\,
simultaneously on one field and 4.5 and 8.0 \um\, on an adjacent
field. Due to this construction and the telescope rotation between the
two epochs, the full area was covered by all bands after completion of
the observations. A summary of the observations is given in Table
\ref{obs.tab}. The raw data and the observational details can be
obtained from the {\it Spitzer} Archive with the Leopard software
package\footnote{http://ssc.spitzer.caltech.edu/propkit/spot/}.
 \begin{deluxetable}{llcc}
     \tablecolumns{2}
     \tablewidth{0pc}
     \tablecaption{Observations\label{obs.tab}
     }
     \tablehead{
       \colhead{} & \colhead{}
     }
     \startdata
     {\it Spitzer} program ID   &   20708           \\
     Target name     &   E-CDFS               \\
     R.A. (J2000)      &   $3^h32^m29.^s46$      \\
     Decl (J2000)     &   $-27\arcdeg 48\arcmin 18\arcsec.32$\\
     Start date ep1  &   2005 Aug 19 (week 91)  \\
     End date ep1    &   2005 Aug 23 (week 91)  \\
     Start date ep2  &   2006 Feb 06 (week 115) \\
     End date ep2    &   2006 Feb 11 (week 116)
     \enddata
   \end{deluxetable}

\section{Data Reduction}\label{reduction}
The reduction of the IRAC data was carried out using the Basic
Calibrated Data (BCD) generated by the {\it Spitzer} Science Center
(SSC) pipeline and a custom-made pipeline that post-processes and
mosaicks the BCD frames. The reduction includes the following steps:\\ 

\begin{my_itemize}
\item SSC pipeline processing (Section\ref{ssc})\\
\item Artifact correction (Section\ref{arts})\\
\item Cosmic ray rejection (Section\ref{cr})\\
\item Astrometry (Section\ref{astr})\\
\item Image combination and mosaicking (Section\ref{mos})\\
\item Flux calibration (Section\ref{zp})\\
\item Exposure time and RMS maps (Section\ref{exp_rms})\\
\item Flag maps (Section\ref{flags})\\
\end{my_itemize}

\subsection{SSC Pipeline Processing}\label{ssc}
The starting point for the reduction are the BCD frames produced by
SSC pipeline. The epoch 1 observations were processed by BCD pipeline
version S12.4.0. The epoch 2 data were processed using pipeline
version S13.2.0. The main differences between these two versions are
related to pointing refinement, muxstriping, and flux conversion. These
issues are all addressed separately in our own reduction pipeline, and
hence these updates have no effect on the end product. An additional
enhancement of S13.2.0 is the introduction of a super sky flat image,
based on the first two years of IRAC of flat-field data. This has only
a small effect on the data of at most 0.5\%. The most significant
steps of the SSC IRAC reduction pipeline are dark subtraction,
detector linearization, flat-fielding, and cosmic ray detection. The
data are calibrated in units of $MJy sr^{-1}$. The pipeline also identifies
bad pixels, which it flags and writes to a mask image, and constructs
initial masks for cosmic rays (called "brmsk").\\
\figc
\figd

\subsection{Post-processing of the BCD Frames}
We post-process the BCD frames to correct for several artifacts caused
by highly exposed pixels (primarily bright stars and cosmic rays) and
calibrate the astrometry. In this section, we briefly describe some of
the artifacts and how we try to remove them. More detailed information
can be found in the IRAC Data Handbook, Section 
4\footnote{http://ssc.spitzer.caltech.edu/irac/dh}. The subsequent
reduction steps are similar, but not identical, to those applied by
the GOODS 
team\footnote{http://ssc.spitzer.caltech.edu/legacy/goodshistory.html}.\\
We start with discarding the two leading short exposures of each
series of observations, which can suffer from the so-called
first-frame effect and cannot be calibrated correctly\footnote{Due to
  the first-frame effect, the first frame of a series of observations
  will have a different bias offset than the rest of the observations
  in the sequence. Since the first image of each series is observed in
  "HDR-mode" (a very short exposure time of 0.4 s for
  identification of saturated sources), the second exposure might
  still suffer from this effect. It is recommended not to include
  these frames when building a mosaic.}.\\
Prior to correction for the artifacts, a median sky image is
constructed based on the data taken in each series of
observations. This sky image is subtracted from each individual frame
to remove both residual structure or gradients in the background
caused by bias or flat fielding, and long-term persistence effects.\\ 

\subsubsection{Detector Artifacts}\label{arts}
One of the principal artifacts in IRAC data is column pulldown. When a
bright star or cosmic ray reaches a level of $> \sim $ 35,000 DN 
in the channel 1 and 2 arrays (3.6 and 4.5 \um), the intensity of the
column in which the bright object lies is affected. Since the
intensity decreases throughout the column, this effect is called
"column pulldown". While column pulldown is slightly different below
and above the bright object and has a small slope, the effect is
nearly constant in practice. We therefore correct for the effect by (1)
locating the columns of $> \sim $ 35,000 DN 
pixels (2) masking all bright sources in the frame, (3) calculating the
median of the affected columns excluding any sources, and (4)
subtracting the median. We favor this simple correction because its
implementation is more robust than fitting, e.g., a general two-segment
slope.\\
Besides column-pulldown, channels 1 and 2 suffer from an effect known
as muxbleed, which appears as a trail of pixels with an enhanced and
additive output level. When a bright source is read out, the readout
multiplexers do not return to their cold state for some time,
resulting in a pattern that trails bright sources on the row. Since
columns are read simultaneously in groups of four, the effect repeats
every fourth column. The amplitude of the effect decreases with
increasing distance to the bright object, but it does not scale with
its flux. It is therefore not possible to fit muxbleed by a simple
function, and we choose for a very straightforward cosmetic
correction. For each offending pixel ($>$ 30 $MJy sr^{-1}$), we generate a
list of pixels selecting every fourth pixel next in the row and
previous in the row. Then, we median filter the pixel list with a
filter width of 20 pixels and subtract the result. The data products
(see Section \ref{images}) include a map that shows which pixels were
muxbleed corrected.\\
This procedure removes the bulk of the muxbleed signal, but not all of
it. However, the effect of a residual muxbleed signal in the final
mosaic is reduced because of the rotation of the field between the two
epochs. At different times, the muxbleed trail affects different pixels
relative to the source position.\\
Bright stars, hot pixels, and particle or radiation hits can also
generate a muxstripe pattern. Where muxbleed only affects pixels on
the same row, the muxstripe pattern may extend over a significant part
of the image, albeit to lower levels. Muxstriping appears as an
extended jailbar pattern preceding and/or following the bright
pixel. It is a fairly subtle effect, usually only slightly visible in
individual frames around very bright stars, but it becomes easily
visible in deeper combined frames. Muxstriping is caused by the
increase of relaxation time of the multiplexer after a bright pixel is
read out. It takes $\sim$10 $\mu$s to clock the next pixel onto an
output, whereas the recovery time after the imprint of a bright pixel
is of the order of tens of seconds. The muxstripe effect also repeats
every fourth column and extends below each source. Each horizontal
band of the image between two bright sources, contains the pattern
induced by all sources above it and needs to be corrected
accordingly.\\
We remove this effect by applying an offset in the zones surrounding
the offending pixels using a program kindly provided by Leonidas
Moustakas of the GOODS-team. In brief, this algorithm identifies the
bright sources in each frame and produces a model of the corresponding
muxstripe pattern, which can then be subtracted. \\
Figures \ref{art1} and \ref{art2} show the treatment of the artifacts
just described. In the left panel, a BCD frame is affected by column
pulldown, muxbleed, and muxstriping. The middle panel shows the
corrections, this frame is subtracted from the affected one which
results in the image on the right, a clean frame.\\
Finally, bright sources leave positive residuals on subsequent
readouts of the array (persistence), although much of the signal
subsides after 6-10 frames. We correct for persistence by creating a
mask of all highly exposed pixels in a frame and then masking those
pixels in the six subsequent frames. Any residual contamination through
persistence will be diminished by the final combination of all
exposures.\\ 
After correction for artifacts, the pipeline subtracts a constant
background by (1) iteratively thresholding and masking pixels
associated with sources and calculating the mode and RMS of the
remaining background pixels and (2) subtracting the mode of the image.

\subsubsection{Cosmic Ray Rejection}\label{cr}
For each series of observations, a first pass registered mosaic is
created from the post-processed BCD frames. For the construction of
this mosaic, the BCD "brmsk"-frames are used as a first guess to mask
candidate cosmic rays. The image is median combined, so it should be
free of any deviant pixels.\\
Next, the first pass image is aligned and subtracted from each
exposure. To create a cosmic ray detection image, the result is
divided by the associated BCD "bunc" image, which contains estimates
of the uncertainties in each pixel based on a noise model\footnote{The
  BCD uncertainty images are the sum of estimates of the read noise,
  the shot noise due to the sky and uncertainties in the dark and flat
  calibration files}. Pixels in this detection image are flagged as
cosmic rays if they deviate more than six times the median value. Pixels
adjacent to deviant pixels are also flagged using a lower threshold
(factor 3.5). These flagged pixels are ignored in the analysis of the
data.\\

\subsubsection{Astrometry}\label{astr}
The SIMPLE astrometry is calibrated to a compact-source catalog 
detected in a combined deep $BVR$ image from MUSYC\footnote{The
  astrometry of the MUSYC $BVR$ detection image is tied to the stellar
  positions of the USNO-B catalog (Monet et al. 2003)} (Gawiser et
al. 2006). The calibration is done on combined frames that were taken
sequentially around the same positions. The combined images are
cross-matched to the $BVR$ source catalog and the positions of the
reference sources are measured.\\
The astrometric differences between the reference catalog and the
SIMPLE pointings are small (up to $\sim$1$\arcsec$) and can be
corrected by applying a simple shift. There is no evidence for
rotation, or higher order geometric distortion. We therefore apply a
simple offset to the WCS CRVAL1 and CRVAL2 of the BCD frames to refine
the pointing. The pointing refinement solutions determined for the 3.6
and 4.5 \um\, BCDs are applied to the 5.8 and 8.0 \um\, images,
respectively, as there are generally few bright sources at 5.8 and 8.0
\um\, to derive them independently. \\
The resulting astrometry accuracy relative to the MUSYC E-CDFS $BVR$
catalog is typically $\sim0\arcsec.09$ (averaged per IRAC channel),
with source-to-source 2-$\sigma$-clipped RMS of $\sim$0$\arcsec$.12 in
channel 1/2 and $\sim$0$\arcsec$.14 in channel 3/4. Large-scale
shears, systematic variations on scales of a few arcminutes, are
0$\arcsec$.2 or less. Figure \ref{astro} shows the residual shifts of
the [3.6 \um] mosaic with respect to the MUSYC $BVR$ image. The quoted
astrometric uncertainties are relative to the MUSYC $BVR$ catalog, but
we also verified that the astrometry agrees very well
($\sim$0$\arcsec$.1 level) with the "wfiRdeep" image (Giavalisco et
al. 2004), which is used as a basis for the ACS GOODS astrometry. \\
\fige
\subsection{Image Combination and Mosaicking}\label{mos}
After individual processing, the individual BCD frames are mosaicked
onto an astrometric reference grid using the refined astrometric
solution in the frame headers.

\subsubsection{Reference Grid}
For the reference grid, we adopt the tangent point, pixel size, and
orientation of the GOODS-IRAC images ($\alpha = 3^h32^m29.^s460,
\delta = -27\arcdeg 48\arcmin 18\arcsec.32$, 0$\arcsec$.6 pixel$^{-1}$. The
pixel axes are aligned with the J2000 celestial axes
\footnote{http://data.spitzer.caltech.edu/popular/goods/20051229\_enhanced \label{goods_ref}}.\\
Also following GOODS, we put the tangent point (CRVAL1,2) at a
half-integer pixel position (CRPIX1,2). This ensures that images with
integer pixel scale ratios (e.g., 0$\arcsec$.3, 0$\arcsec$.6,
1$\arcsec$.2) can (in principle) be directly rebinned (block summed or
replicated) into pixel alignment with one another. This puts GOODS,
SIMPLE, and FIDEL (a deep 24/70 \um\, survey in the E-CDFS) on the same astrometric
grid. The final SIMPLE mosaic extends 38\arcmin \, $\times$ 48\arcmin \,
(3876 $\times$ 4868 pixels).\\

\subsubsection{Image Combination}
For each epoch, the individual post-processed BCD frames are
transformed to the reference grid using bicubic interpolation, taking
into account geometric distortion of the BCD frame. Cosmic rays and
bad pixels are masked and the frames are average combined without
additional rejection.\\
Finally, the separate epoch 1 and epoch 2 mosaics are combined,
weighted by their exposure times. By design, the SIMPLE E-CDFS
observational strategy maps around the GOODS-S field, which leaves a
hole in the combined mosaic. To facilitate the analysis, we add the
GOODS-S IRAC data (DR3, mosaic version 0.3 $^{\ref{goods_ref}}$, to
the center of the SIMPLE mosaic. We shift the GOODS-S IRAC mosaics by
$\sim$0$\arcsec$.2 to bring its astrometry in better agreement with
SIMPLE. To ensure a seamless combination between the epoch 1, epoch 2,
and GOODS-S images, we subtract an additional background from the
images before combination. The background algorithm masks sources and
measures the mode of the background in tiles of 1\arcmin \, $\times$
1\arcmin. The "mode map" is then smoothed on scales of 3\arcmin \,
$\times$ 3\arcmin \, and subtracted from the image, resulting in
extremely flat images and a zero background level on scales $>$
1\arcmin.\\
 \begin{deluxetable*}{lccccc}
     \tablecolumns{6}
     \tablewidth{0pc}
     \tablecaption{\label{conv.tab}
     }
     \tablehead{
       \colhead{channel} & \colhead{$\lambda$} & \colhead{flux
         conversion\tablenotemark{a}} &  \colhead{zeropoint} &
       \colhead{FWHM} & \colhead{Gaussian}\\
       & && & & convolution \\
       & (\um) & ($\mu$Jy (DN s${-1}$)$^{-1}$)& (AB) & (\arcsec) & (\arcsec)\\
     }
     \startdata
     ch1   & 3.6  &  3.922    &   22.416  &   1.97  &  0.84  \\
     ch2   & 4.5  &  4.808    &   22.195  &   1.93  &  0.93  \\
     ch3   & 5.8  & 20.833    &   20.603  &   2.06  &  0.80  \\
     ch4   & 8.0  &  7.042    &   21.781  &   2.23  &   --
     \enddata
     \tablenotetext{a}{-Listed as FLUXCONV in the image headers}
     \tablecomments{The FWHM of the $U-K$ images is 1\arcsec.5. To convolve
       those to the PSF of ch4, we use a sigma of 1.34.}
     
   \end{deluxetable*}

\subsection{Flux Calibration}\label{zp}
The SSC data are calibrated using aperture photometry in 12\arcsec\,
apertures. Since ground-based IR calibrators are too bright to use for
IRAC, the actual flux for each channel needs to be predicted using
models (Cohen et al. 2003). The resulting calibration factors were
determined by Reach et al. (2005) and are listed in the image headers
and Table \ref{conv.tab}.\\
The epoch 1 and epoch 2 science images were scaled to a common
zeropoint so that their data units agree. For convenience, we
calibrate our images to the GOODS-S IRAC data (in DN s$^{-1}$). This
is done using the original calibration factors from Table
\ref{conv.tab}. The relative accuracy of the zeropoint can be
estimated by minimizing the count rate differences of bright,
non-saturated stars in circular apertures in regions where the images
overlap. This indicates that the fluxes agree within $\sim$3\%.

\figg
\figf
\subsection{Additional Data Products}
\subsubsection{Exposure Time and RMS Maps}\label{exp_rms}
The exposure time maps are created by multiplying, at each position,
the number of BCD frames that were used to form the final image by the
integration time of each frame. The exposure map thus reflects the
exposure time in seconds on that position of the sky, not the average
exposure time per final output pixel.\\
The 25\%, 50\%, and 75\% percentiles of the final exposure maps
(excluding GOODS-S) are $\sim$3100, 5500 and 9100 s (0.9, 1.5 and
2.5 hr) for all channels. The corresponding area with at least that
exposure time is $\sim$1200, 800, and 400 arcmin$^2$,
respectively. In addition, the central GOODS-S mosaic has $\sim$23
hours per pointing over $\sim$138 arcmin$^2$.\\
This release also provides RMS maps. The RMS maps were created by (1)
multiplying the final mosaic by the $\sqrt{(t_{exp}/median(t_{exp}))}$
(where $t_{exp}$ is the exposure time map), to create an exposure
normalized image, (2) iteratively rejecting pixels deviating $> 4.5\,
\sigma$ and their directly neighboring pixels, (3) binning the image
by a factor 4 $\times $ 4, and (4) calculating the RMS statistic of the
binned pixels in a moving window of 15 $\times$ 15 bins. The result is
approximately the local RMS background variation at scales of
2\arcsec.4\, at the median exposure time, which does not suffer from
correlations due to resampling. We multiply this value by
$\sqrt{4}/\sqrt{(t_{exp}/median(t_{exp}))}$ to get the approximate
per-pixel RMS variation at the mosaic pixel scale for other exposure
times (see, e.g., Labb\'e et al. 2003). This RMS map does not directly
reflect the contribution to the uncertainty of source confusion. The
variations in the RMS due to instrumental effects are mitigated by the
addition of the observed epochs under 180\degree\, different roll
angles. 

\subsubsection{Flags}\label{flags}
We provide a flag map, which identifies pixels corrected for muxbleed
in channel 1 and channel 2. These corrections are not optimal, and
when analyzing the images or constructing source catalogs, it may be
useful to find pixels which may have been affected. The flag image is
a bit map, i.e., an integer map that represents the sum of bit-wise
added values (flag = 1 indicates a muxbleed correction in the first
epoch, flag = 2 indicates a correction in the second epoch).

\subsection{Final Images}\label{images}
The final images of SIMPLE are publicly
available\footnote{http://data.spitzer.caltech.edu/popular/simple}. The
data release consists of FITS images of all IRAC observations in the
E-CDFS. We provide science images, exposure time maps, RMS maps, and a
flag map. These images comprise combined mosaics of all data taken
(both epochs), including the 10\arcmin \, $\times$ 15\arcmin \, GOODS
IRAC mosaics in the center. In addition, we provide combined mosaics
and exposure maps of the data of the individual epochs (without the
GOODS data), which may be useful to study the reliability and/or
variability of sources. The units of the science and RMS images are DN
s$^{-1}$, with the (GOODS) zeropoints as given in Table
\ref{conv.tab}. The units of the exposure time maps are
seconds. Figures \ref{colorz} and \ref{color} show the color composite
image of the 3.6 \um\, and 5.8 \um\, mosaics.\\

\section{Additional Data}\label{ancil}
\subsection{The $U-K$ Data}
To cover the optical to NIR regime, we use the $UBVRI$ imaging from
the COMBO-17 and ESO DPS surveys (Wolf et al. 2004 and Arnouts et
al. 2001, respectively) in the re-reduced version of the GaBoDS
consortium (Erben et al. 2005; Hildebrandt et al. 2006). We include
the $z'JHK$ images from the Multiwavelength Survey by Yale-Chile
(MUSYC, Gawiser et al. 2006), which are available
online\footnote{http://www.astro.yale.edu/MUSYC}. The final
$UBVRIz'JHK$ images typically have a seeing of $\sim 1\arcsec$.  The
images we use were PSF-matched to the image with the worst seeing
($J$-band, 1\arcsec.5) by Taylor et al. (2009b). For more details on the
construction of the MUSYC survey and the different data sets, we refer
the reader to Taylor et al. (2009b).
 
\subsection{The MIPS 24 \um\, Data}
The E-CDFS was also observed extensively by the Multi-band Imaging
Photometer for {\it Spitzer} (MIPS) as part of FIDEL (PI:
M. Dickinson). The survey contains images at 24, 70, and 160 \um. We
only consider the 24 \um\, image, due to its utility as an indicator of
star formation, its sensitivity, and the fact that the source
confusion at 24 \um\, is less severe compared to the longer
wavelengths. The FIDEL 24 \um\, image reaches a 5-$\sigma$ sensitivity
ranging from 40 to 70 $\mu$Jy, depending on the source position
(Magnelli et al. 2009). We use the v0.2 mosaic, which was released on
a scale of 1\arcsec.2\, pixel$^{-1}$. 

\section{Source Detection and Photometry}\label{detphot}
\subsection{Detection}\label{det}
Sources are detected and extracted using the SExtractor software
(Bertin \& Arnouts 1996) on a detection image. The detection image is
an inverse-variance weighted average of the 3.6 and 4.5 \um\,
images. The 3.6 and 4.5 \um\, bands are the most sensitive IRAC bands
and the combination of the two leads to a very deep detection
image. To enable detection to a similar signal-to-noise limit over the
entire field, we multiply the [3.6]+[4.5] image by the square root of
the combined exposure map. This produces a "noise-equalized" image
with approximately constant signal to noise, but different depth, over
the entire field. Figure \ref{fov} shows the noise-equalized detection
image in the background.\\
Subsequently, we run SExtractor on the detection map with a 2-$\sigma$
detection threshold. We choose this detection limit to be as complete
as possible, at risk of severe confusion. We will discuss the matter
of confusion later. In the detection process, SExtractor first
convolves the detection map with a detection kernel optimized for
point sources. We use a 5 $\times$ 5 convolution mask of a Gaussian PSF with an
FWHM of 3 pixels. Furthermore, we require a minimum of two adjacent
pixels above the detection threshold to trigger a detection. The
resulting catalog contains 61,233 sources, 43,782 of which have a
signal-to-noise ratio (S/N) $> 5$ at 3.6 \um.\\
Instead of our exposure time-detection image, we could have used the
RMS map for detection. In practice, the RMS should be proportional to
$1/\sqrt{(t_{exp})}$ and the choice of detection image should not
significantly influence the output catalog. To test the correspondence
of RMS and $1/\sqrt{(t_{exp})}$, we multiplied the RMS by the square
root of the exposure time map, which results in a tight Gaussian
distribution with a width of $\sigma$ = 0.003. Our exposure time
detection image is therefore very similar to a detection image based
on an RMS map.\\ 
As an aside, we note that SExtractor's RMS map underestimates the true
noise as the pixels are correlated (see, e.g., Labb\'e et al. 2003). If
we use SExtractor's RMS map, we find $\sim$10\,\% more objects than with
our method, as expected. Many of these objects are near the edges of
the image; none of them have an S/N $>$ 5. 

\subsection{Photometry}\label{phot}
\subsubsection{Image Quality and PSF Matching}\label{psf}
In order to obtain consistent photometry in all bands, we smooth all
images (except MIPS) to a common PSF, corresponding to that of the 8.0
\um\,, which has the broadest FWHM. To determine the FWHM, we compile a
list of stars with $(J-K_s) < 0.04$. We select five different
areas of the E-CDFS to check whether the PSF changes over the
field. This is in particular important for the IRAC bands, which have
a triangular-shaped PSF. Because of the rotation between the two
epochs, the final IRAC PSF is a combination of two triangular-shaped
PSFs that are rotated with respect to each other. This combined PSF
can vary with position in the field of view and we first need to check
how large these variations are. Radial profiles of the stars are
determined using the IRAF task {\tt imexam}.
We find that the variation of the mean FWHM over the whole field of
view is $ <$ 5\% for all IRAC bands and there is no clear trend
between the mean FWHM and the position on the field for any IRAC
band. We convolve all images with a Gaussian to produce similar PSFs
in all bands. The mean original FWHM per band and the Gaussian sigma
values used for convolution are listed in Table \ref{conv.tab}.

\subsubsection{$UBVRIz'JHK$ + IRAC}
We run SExtractor in dual-image mode, meaning that the program
determines the location of sources in the combined [3.6]+[4.5]
detection image, and then measures the fluxes in the smoothed science
images in the exact same apertures. We perform photometry in fixed
circular aperture measurements in all bands for each object, at radii
of 1$\arcsec$.5, 2$\arcsec$.0, and 3$\arcsec$.0. In addition, we use
SExtractor's autoscaling apertures based on Kron (1980)
radii. Following Labb\'e et al. (2003), we refer to these apertures as
APER(1.5), APER(2.0), APER(3.0), and APER(AUTO). We use these
apertures to derive both color fluxes and total fluxes (see Labb\'e et
al. 2003).\\
SExtractor provides a flag to identify blended sources that we include
in our catalog as ``flag\_blended''. In the SIMPLE catalog,
$\gtrsim$60\%\footnote{Sixty-two percent of the sources suffer from blending
  (SExtractor's FLAGS keyword=1), 61\% of the sources have a close
  neighbor (FLAGS = 2), and for 66\% of the sources FLAGS=1 $\vee$
  FLAGS=2.} of all sources are flagged as blended. This is due to the
large PSF of the camera and the depth of the image. \\
Given the large number of blended sources, it is useful to be able to
identify only the most extreme cases of blending. If the sum of the
Kron aperture radii of a source and its nearest neighbor exceeds their
separating distance and if the neighbor's flux is brighter than its
own, we set the 'flag\_blended' entry to 4. The percentage of sources
suffering from this form of extreme blending is 32\% for all sources
with S/N$ > 5$ at 3.6 \um.\\
While performing photometry, we treat blended sources
separately. Following Labb\'e et al (2003) and Wuyts et al. (2008), we
use the flux in the color aperture to derive the total flux for
sources that suffer from severe blending. For the identification of
blended sources, we prefer our own conservative blending criterion over
SExtractor's blending flag, since the latter identifies too many
sources as blended, even sources for which the photometry is
essentially unaffected \footnote {Adopting the SExtractor blending
  flag would produce a catalog that mostly consists of blended sources
  ( $\sim$90\% for sources with a $5-\sigma$ detection at 4.5 \um\,
  and in the $K$-band). These would all be assigned color fluxes that
  are, in our case, measured within a fixed aperture. The effect such
  a large fraction of aperture fluxes has on the comparison with the
  MUSYC catalog can be seen in Fig.~\ref{fluxapers} of Appendix
  \ref{aper.app}. The upper left panel shows a large tail of bright
  sources that are significantly offset with respect to a one-to-one
  relation.}. If we do not make a distinction between blended and
non-blended sources, the comparison with other catalogs worsens
slightly ($ < $ 0.02 mag on the mean deviation).\\
To determine the color fluxes, we use the circular apertures with
2\arcsec\, radius for all sources in all bands:

\begin {equation}
{\small
APER(COLOR)=  APER(2.0).
}
\label{col_ap.eq}
\end {equation}

We calculate the total fluxes from the flux measured in the AUTO
aperture. For sources with an aperture diameter smaller than
4\arcsec\, diameter, we apply a fixed aperture of 4\arcsec.

\begin {equation}
{\small
APER(TOTAL)=\left\{\begin{array}{cl}
	       \vspace{0.2cm}
               APER(AUTO),  &   Ap_{tot} > 4\arcsec  \\
	       APER(COLOR), &   Ap_{tot} \leq 4\arcsec 
            \end{array}\right.
}
\label{tot_ap.eq}
\end {equation}

Where $Ap_{tot}$ is the circularized diameter of the Kron aperture. If
the source is blended (FLAG\_BLENDED = 4), then 
\begin
  {equation}{\small APER(TOTAL)=APER(COLOR)}\nonumber  
\end {equation}
Finally, we apply an aperture correction to the total fluxes using the
growth curve of bright stars to correct for the minimal flux lost
because it fell outside the "total" aperture.\\
For the IRAC data, we apply individual growth curves for each band. The
zeropoint for the aperture correction is based on the values listed in
Table 5.7 of the IRAC Data
Handbook\footnote{http://ssc.spitzer.caltech.edu/irac/dh}. We use the
zeropoint in an aperture of 7\arcsec.3 diameter (3 pixel radius in
Table 5.7)\footnote{We use this aperture instead of the more generally
  used 12\arcsec\, diameter because of the high density of sources in
  our field, which would lead to source confusion at large radii. To
  avoid these complications, we determine the inner part of the growth
  curve from our data to a 3\arcsec.66 radius, and combine it with the
  tabulated values from the handbook at larger radii. In this way, we
  minimize the effect of blending.}. For the MUSYC optical-IR data, we
use the $K$-band growth curve to correct the total fluxes in all
bands. The aperture corrections are listed in Table \ref{apcor.tab}.
  \begin{deluxetable}{lccc}
     \tablecolumns{4}
     \tablewidth{0pc}
     \tablecaption{Aperture
       corrections\tablenotemark{a}\label{apcor.tab}
     }
     \tablehead{
       \colhead{band} & \colhead{4\arcsec-7\arcsec.3} &
       \colhead{7\arcsec.3-12\arcsec} & \colhead{total correction}
     }
     \startdata
     $K$       &       --   &    --      &  1.28 \\
     3.6 \um   &  1.22      &   1.112    &  1.35  \\
     4.5 \um   &  1.24      &   1.113    &  1.38  \\
     5.8 \um   &  1.37      &   1.125    &  1.55  \\
     8.0 \um   &  1.42      &   1.218    &  1.73  \\
     \enddata
     \tablenotetext{a}{Expressed as total flux divided by the flux in
       APER(COLOR)}
     \tablecomments{4\arcsec\, is our color aperture, 7\arcsec.3 is
       taken from Table 5.7 from the IRAC Data Handbook (corresponds
       to 3 pixel radius in that table), and 12\arcsec\, is the
       zeropoint aperture (see Section~\ref{zp}). The numbers in the
       second column are derived from our growth curves, the third
       column contains the corrections from the Data Handbook, and the
       total corrections are listed in the last column.} 
   \end{deluxetable}
 \figh
\figi
\subsubsection{The MIPS 24 \um\, Data}\label{mips}
The photometry of the MIPS 24 \um\, image is performed in a different
way, because of the larger PSF. Here, we use a deblending model to
mitigate the effects of confusion. We use the source positions of the
IRAC 3.6 \um\, image, which has a smaller PSF, to subtract modeled
sources from MIPS sources that show close neighbors, thus cleaning the
image. After this procedure we perform aperture photometry in
apertures of 6\arcsec\, diameter, and correct fluxes to total fluxes
using the published values in Table 3.12 of the MIPS Data Handbook.\\
In principle a similar approach could have been attempted for the IRAC
images themselves. Ground-based $K$-band data and space-based NICMOS
imaging have been successfully used to deblend IRAC images (Labb\'e et
al. 2006, Wuyts et al. 2008). However, whereas the resolution of our
$K$-band image is appropriately high, the image is not deep enough for
this kind of modeling.

\subsection{Background and Limiting Depths}\label{noise}
The determination of the limiting depth depends on the noise
properties of the images. To analyze those, we place $\sim$4000
circular apertures on the registered and convolved images and measure
the total flux inside the apertures. Apertures are placed across the
field in a random way, excluding all positions associated with sources
using the SExtractor segmentation map. We use identical aperture
positions for all bands, and repeat the measurements for different
aperture sizes. The distribution of empty aperture fluxes can be
fitted by a Gaussian, which provides the flux dispersion of the
distribution. The RMS depends on aperture size and is larger for
larger apertures (see Fig.\,\ref{rms}). The left panel shows the
distribution of empty aperture fluxes for channel 1 for apertures of
sizes 2\arcsec, 3\arcsec, and 4\arcsec. The right panel shows how the
RMS increases with aperture size for all IRAC bands. The noise level
is higher than can be expected from uncorrelated Gaussian noise. The
reason for this is that correlations between neighboring pixels were
introduced while performing the data reduction and PSF matching (see
also Labb\'e et al. 2003). \\
The depth of our SIMPLE IRAC mosaic is a function of position, as some
parts have longer exposure times than others. Table \ref{depth.tab}
lists the total AB magnitude depths at 5-$\sigma$ for point sources
and the area over which this depth is achieved. Figure~\ref{mag_lim}
provides a graphic representation of the limiting depths of all
wavelength bands.\\
\figj
To investigate whether our measurement of the uncertainties in the
IRAC photometry is reasonable, we compare the IRAC fluxes of epoch 1
with those of epoch 2.
 The results are shown in
Fig.\,\ref{ep1ep2}. The median offsets between the two epochs are
printed in the lower left corner and are close to zero in each
band. The scatter in each panel is small and comparable to the
estimated RMS values. \\
  \begin{deluxetable}{lcccl}
     \tablecolumns{5}
     \tablewidth{0pc}
     \tablecaption{5-$\sigma$ limiting depths (total AB
       magnitude)\label{depth.tab}
     }
     \tablehead{
       \colhead{} & \colhead{}
     }
     \startdata
     percentile & 75\%       & 50\%   & 25\%   & (percentile of
     pixels)\\
     exptime    & $>$0.9     & $>$1.5 & $>$2.5 & (hours) \\
     area       & $\sim$1200 & 800    & 400    & (area in arcmin$^2$
     with at \\
                &            &        &        &  least this exposure
                time) \\
     3.6 \um    & 23.66      & 23.86  & 24.00  & (depth at 3.6 \um)\\
     4.5 \um    & 23.50      & 23.69  & 23.82  & (depth at 4.5 \um)\\
     5.8 \um    & 21.68      & 21.95  & 22.09  & (depth at 5.8 \um)\\
     8.0 \um    & 21.69      & 21.84  & 21.98  & (depth at 8.0 \um)\\
     \enddata
   \end{deluxetable}

\subsection{Stars}\label{stars}
We identify stars by their color and signal to noise ($J - K < 0.04\,
\wedge\, wK > 0.5\, \wedge\, (S/N)_K > 5$) and find 978 stars in the
total catalog. To test the validity of this selection criterion, we
compare it to the $BzK$ selection technique defined by Daddi et
al. (2005). In the $BzK$ diagram stars have colors that are clearly
separated from the colors of galaxies and they can be identified with
the requirement $(z - K) < 0.3 \cdot (B - z) - 0.5$. From the 978
stars in the SIMPLE catalog with sufficient signal to noise in the
$B$- and $z$-bands, 94\% obey the $BzK$ criterion. In the
$BzK$ diagram, the remaining 6\% lie only slightly above the
$BzK$ stellar selection limit.

\section{Derived Parameters}\label{derpar}
\subsection{Spectroscopic and Photometric Redshifts}\label{redshifts}
The E-CDFS is one of the principal fields for high-redshift studies
and has consequently been the object of many spectroscopic
surveys. Taylor et al. (2009b) compiled a list of reliable 
spectroscopic redshifts from several of these surveys, which we
cross-correlated with our SIMPLE catalog. The spectroscopic redshifts
come from: Croom et al. (2001), Cimatti et al. (2002), le F\`evre et
al. (2004), Strolger et al. (2004), Szokoly et al. (2004), van der Wel
et al. (2004, 2005), Daddi et al. (2005), Doherty et al. (2005),
Mignoli et al. (2005), Ravikumar et al. (2007), Kriek et al. (2008),
Vanzella et al. (2008), Popesso et al. (2009), and Treister et
al. (2009a). The list contains 2095 spectroscopic redshifts. \\
In addition, we include photometric redshifts from the COMBO-17 survey
(Wolf et al. 2004) out to $z$= 0.7, which are very reliable at those
redshifts. For the remainder of the sources we compute photometric
redshifts using the photometric redshift code EAZY (Brammer et
al. 2008). The EAZY algorithm provides a parameter $Q_z$ that
indicates whether a derived photometric redshift is reliable. Brammer
et al. (2008) show that for $Q_z > 2-3$ the difference between
photometric and spectroscopic redshifts increases sharply and that
quality cuts based on $Q_z$ can reduce the fraction of outliers
significantly. Therefore, when testing the accuracy of our photometric
redshifts, we only include sources with $Q_z < 2$. \\
\figk
Figure \ref{photz} shows the EAZY photometric redshifts compared
against a list of spectroscopic redshifts. The upper panel shows the
direct comparison for sources with $S/N > 5$ in both $K$band and 3.6
\um\, (a total of 1280 sources, from which we remove 54 sources with
$Q_z \ge 2$ (4\%), resulting in a final sample of 1226 sources). The
lower panel shows $\Delta z/(1+z_{spec})$, where $\Delta z = z_{\rm
  phot} - z_{\rm spec}$. X-ray detections are shown in gray. \\
To quantify the scatter, we determine the normalized median absolute
deviation ($\sigma_{NMAD} = 1.48 \times \textrm{median } | x -
\textrm{median}(x)|$, which is a robust estimator of the scatter,
normalized to give the standard deviation for a Gaussian
distribution). Overall the $\sigma_{NMAD}$ of $|\Delta z|/
(1+z_{spec})$ is 0.025, but it varies with redshift, ranging from
0.024 at $z\sim 1$, 0.055 at $z\sim 1.5$, and 0.38 at $z\sim$
2.0. There is a significant fraction (8.2\%) of outliers with $|\Delta
z|/ (1+z_{spec}) > 5 \sigma_{NMAD}$. This number agrees well with the
11\% Taylor et al. (2009b) found for the MUSYC catalog. Many of the
outliers are detected in X-ray and are AGN candidates (43\%). The high
fraction of (candidate) AGN outliers could be explained by the fact
that we do not have an AGN spectrum in our template set. EAZY
photometric redshifts for X-ray detections are, therefore,
uncertain. If we remove them from the sample, the overall accuracy
improves and $\sigma_{NMAD}$ becomes 0.024, 0.041, and 0.16 at
redshifts $z\sim 1.0$, 1.5, and 2.0, respectively. \\ 
Including AGN templates will improve the overall accuracy of the
photometric redshift, as can be seen in Luo et al. (2010). Those
authors determine the photometric redshifts of a sample of X-ray
sources in the E-CDFS, using UV-to-IR data, including the SIMPLE
3.6\um-band. Apart from standard galactic templates they used 10
different AGN templates. In the comparison with spectroscopic
redshifts, there are only three outliers, which is 1.4\% of the sample and
the $\sigma_{NMAD} =$ 0.010.\\ 
We also check whether the outliers suffer from blending. Out of the
101 outliers, 26 sources have a neighboring source whose APER(AUTO)
exceeds their separating distance and whose flux is at least as bright
as its own, which can affect their photometry. However, removing these
sources from the sample does not decrease $\sigma_{NMAD}$, since there
are many sources with nearby bright companions whose photometric and
spectroscopic redshifts agree well.\\

\subsection{Star Formation Rates, Rest-frame Photometry and Stellar
  Masses}
In this section, we describe the main characteristics of the 
procedures for deriving star formation rates and stellar masses. For a
more detailed description, the reader is referred to Damen et
al. (2009a). We estimated SFRs using the UV and IR emission of the
sample galaxies. We use IR template spectral energy distributions
(SEDs) of star forming galaxies of Dale \& Helou (2002) to translate
the observed 24 \um \, flux to $L_{IR}$. First, we convert the
observed 24 \um \, flux density to a rest-frame \lum \,density at
$24/(1+z) $\,\um, then we extrapolate this value to a total IR \lum \,
using the template SEDs. To convert the UV and IR \lums \, to an SFR,
we use the calibration from Bell et al. (2005), which is in
accordance with Papovich et al. (2006), using a Kroupa IMF:

\begin{equation}\label{eqn:ir_sfr}
\Psi / \msol \,\,\mathrm{yr}^{-1} = 1.09 \times 10^{-10} \times
(L_\mathrm{IR} + 3.3\,\, L_{2800}) / \lsol,
\end{equation}\\

where $L_{2800} = \nu L_{\nu,(2800 \AA)}$ is the \lum \, at rest frame
2800 \AA, a rough estimate of the total integrated UV \lum
\,(1216-3000\AA). \\
To obtain stellar masses, we fitted the UV-to-8 \um \, SEDs of the
galaxies using the evolutionary synthesis code developed by Bruzual \&
Charlot 2003. We assumed solar metallicity, a Salpeter IMF, and a
Calzetti reddening law. We used the publicly available HYPERZ stellar
population fitting code (Bolzonella et al. 2000) and let it choose
from three star formation histories: a single stellar population (SSP)
without dust, a constant star formation (CSF) history, and an
exponentially declining star formation history with a characteristic
timescale of 300 Myr ($\tau 300$), the latter two with varying amounts
of dust. The derived masses were subsequently converted to a Kroupa
IMF by subtracting a factor of 0.2 dex. We calculated rest-frame
luminosities and colors by interpolating between observed bands using
the best-fit templates as a guide (see Rudnick et al. 2003 and Taylor
et al. 2009a for a detailed description of this approach and an IDL
implementation of the technique dubbed
'InterRest'\footnote{http://www.strw.leidenuniv.nl/$\sim$ent/InterRest}).\\

\section{Catalog Contents}\label{catalog}
The SIMPLE IRAC-selected catalog with full photometry and explanation
is publicly available on the
Web\footnote{http://www.strw.leidenuniv.nl/$\sim$damen/SIMPLE\_release.html}. We
describe the catalog entries below.
\figl
\figm
\begin{itemize}
\item { $ID$}              --- A running identification number in
  catalog order as reported by SExtractor.
\item { $x\_pos,y\_pos$}   --- The pixel positions of the objects
  based on the combined 3.6 \um\, + 4.5 \um\, detection map.
\item { $ra, dec$}         --- The right ascension and declination in
  equinox J2000.0 coordinates, expressed in decimal degrees.
\item { $i\_colf $}        --- Observed color flux in bandpass $i$,
  where $i = U, B, V, R, I, z', J, H, K, irac1,$ $irac2, irac3, irac4$
  in circular apertures of 4\arcsec\, diameter. All fluxes are
  normalized to an AB magnitude zeropoint of 25. 
\item { $i\_colfe $}       --- Uncertainty in color flux in band $i$
  (for derivation see Section \ref{noise}).
\item { $j\_totf$}         --- Estimate of the total flux in band $j$,
  where $\{j = K, irac1, irac2, irac3, irac4\}$, corrected for missing
  flux assuming a PSF profile outside the aperture, as described in
  Section \ref{psf}.
\item { $j\_totfe$}        --- Uncertainty in total flux in band $j$.
\item { $ap\_tot\_j$}      --- Aperture diameter (in \arcsec) used for
  measuring the total flux in band $j$. This corresponds the
  circularized diameter of APER(AUTO) when the Kron aperture is
  used. If the circularized diameter is smaller than 4\arcsec, the
  entry is set to APER(COL) = 4\arcsec\, (see Section \ref{phot}). 
\item { $iw$}              --- Relative weight for each band $i$. For
  the IRAC bands, the weights are determined with respect to the
  deepest area of the SIMPLE mosaic without GOODS.
\item { $flag\_star$}      --- Set to 1 if the source meets the
  criteria of Section \ref{stars}.
\item { $flag\_blended$}   --- Contains the SExtractor deblending
  flag, which indicates whether a source suffers from blending (bit =
  1) or whether it has a close neighbor (bit = 2). If a source suffers
  from extreme blending (see Section \ref{phot}) then bit = 4.
\item {$flag\_qual$}       --- Bitwise added quality flag, that
  indicates whether a source lies in the GOODS area (bit = 1), lies in
  a stellar trail (bit = 2), falls outside the MUSYC field (bit = 4)
  or has been corrected for muxbleed.
\end{itemize}

Please note that all flux units in the catalog are converted to the
same zeropoint on the AB system: $AB\_MAG = 25. - 2.5\log(flux)$. \\

\section{Comparison to Other Catalogs}\label{comp}
In this section, we compare our SIMPLE catalog to the published
catalogs of Taylor et al. (2009; MUSYC, E-CDFS) and  Wuyts et
al. (2008; FIREWORKS, CDFS). All catalogs cover (parts of) the same
area in the sky. The important difference is that we detect sources in
the IRAC 3.6 \um\, and 4.5 \um\, bands, whereas both the MUSYC and the
FIREWORKS catalogs are $K$-band detected. The advantage of an
IRAC-selected catalog is that IRAC probes the rest-frame NIR out to
high redshift. The downside of IRAC selection is the lower resolution,
which leads to confusion. The FIREWORKS catalog used a $K$-band
selection specifically for this reason. We will investigate the effect
these differences have on the catalogs below.
\figii
\subsection{SIMPLE versus MUSYC}\label{comp_musyc}
The optical-NIR part of the SIMPLE catalog ($U-K$) is based on the same
data as the MUSYC catalog. The differences lie in the PSF, detection
method, and photometry. Taylor et al. (2009b) determine their total
fluxes in a similar way as we do. However, they include an extra
correction based on the measurement of the background, which they
measure themselves instead of using the value derived by
SExtractor and they do not make a distinction between blended and
non-blended sources. We cross-correlated the two catalogs and in
Fig. \ref{simple_musyc} we present the comparison. Each panel shows
sources with S/N $>$ 10 in IRAC 4.5 \um\, and in the relevant band of
the panel. We also applied a weight cut in $K$, $wK > 0.75$,
recommended by Taylor et al. (2009b). We determined the median offsets
in different magnitude bins and show them at the bottom of each
panel. The first number ({\it in black}) represents the median offset
of all sources, the gray numbers represent the median offset in each
magnitude bin; they are $\lesssim 0.05$ in all bands. The error bars
represent the formal expected photometric errors, which are dominated
by the Poisson uncertainties in the background. The offsets at bright
magnitudes are not caused by Poisson statistics, but most likely by
slight systematic differences in methodology. We investigated the
bright sources in the $U$-, $B$-, $V$-, and $R$-band, which show an
offset of $>$ 0.2 in color and found that this is an effect
of the aperture sizes that were used. The MUSYC fluxes were determined
using SExtractor's MAG\_ISO, enforcing a minimum aperture diameter of
2\arcsec.5. For the SIMPLE catalog, we used a fixed 4\arcsec.0 aperture
diameter. The large color differences at the bright end occur for
galaxies for which the differences in aperture size are large too
(factor 1.5 and greater).

\subsection{SIMPLE versus FIREWORKS}\label{comp_fw}
\subsubsection{Photometry}
The FIREWORKS catalog is constructed from observations in wavelength
bands that in some cases differ from the ones we use. The $UBVR$ and $I$
data come from the Wide Field Imager and are the same as we use,
except for the $U$-band, for which the FIREWORKS uses the
$U_{38}$-imaging. The $z_{850}$-band image was observed by HST, $J,
H$, and $K_s$ data come from ISAAC. The IRAC images were taken by the
GOODS team and are nearly the same as the ones we use. Figure
\ref{simple_fw} shows the comparison of all these bands against each
other. As in Fig. \ref{simple_musyc}, we only show sources with S/N
$>$ 10 in IRAC 4.5 \um\, and in the relevant band of the panel, with a
weight in $K$-band larger than 0.5. The median values are once more
shown at the bottom left and the error bars again represent the
expected formal errors. \\
The FIREWORKS catalog allows easy identification of blended sources
and we have removed these from Fig.\,\ref{simple_fw}, since they
worsened the comparison. This can be seen in Fig.\, \ref{kblend} in
Appendix \ref{conf.app}, which shows the difference in $K$-band
magnitude for FIREWORKS and SIMPLE. In that figure, we did include the
blended FIREWORKS sources and marked them in red. They form a specific
tail and we have removed them from all further analysis. The sources
that suffer from extreme blending in the SIMPLE catalog do not take
up such a specific locus in the comparison figures. Excluding them
from the sample does not significantly affect the comparison and
therefore we keep them in the sample (see Appendix \ref{conf.app}).\\
In Fig.\,\ref{simple_fw}, the comparison between FIREWORKS and SIMPLE
tails upward at the faint end. There, the SIMPLE fluxes are brighter
than FIREWORKS. This could be due to the fact that the SIMPLE
apertures are quite large and will catch some light from neighboring
sources. \\

A direct comparison between SIMPLE and FIREWORKS illustrates the
strengths of both data sets as can be seen in Fig. \ref{col_mag}, which
shows a color-magnitude diagram of both catalogs for sources with S/N
$>$ 5 in the relevant bands. The envelopes at the bright end agree
well, but at the faint end FIREWORKS reaches greater depth. The
advantage of the SIMPLE survey is its large area, and thus its large
number of sources. Out to a magnitude of 21.5 in [3.6], the SIMPLE
catalog contains 4061 sources at 5-$\sigma$, compared to 1,250 for
FIREWORKS.
\figo
\subsubsection{Derived Properties}\label{derived}
In addition to a comparison of the photometry, we compare derived
quantities of the FIREWORKS and SIMPLE catalogs. Figure \ref{fw_other}
shows the comparison between mass, (specific) star formation rate,
MIPS 24 \um\, flux, and redshift. Mean values in bins of equal number
of sources are indicated by the red line and given at the bottom of
each panel. \\
The panels with MIPS 24 \um\, flux and SFR show the best agreement,
although the scatter in the comparison of the SFR is substantially
higher than it is for the MIPS fluxes. This is caused by the
difference in photometric redshifts. If we use FIREWORKS photometric
redshifts to determine the SIMPLE SFRs, the scatter in the SFRs is
reduced to the scatter in MIPS fluxes.\\
The scatter is highest in the panels where masses and specific star formation rates (sSFRs) are
compared, quantities that depend on photometric redshifts and model
assumptions. These are, therefore, more susceptible to systematic
errors. Since the masses are derived in similar ways for SIMPLE and
FIREWORKS (same models, dust extinction law, metallicity, and IMF),
systematics in the modeling can not be responsible in this comparison. We
redetermined our masses using FIREWORKS photometric redshifts and
found that this reduces the number of outliers in the mass-comparison
panel, but not the scatter. The main reason for the scatter in mass and
sSFR is signal to noise. The mean absolute deviation of the scatter in
the mass comparison is 0.5 for sources with (S/N)$_K <$ 10. For
sources with a (S/N)$_K \sim$20, the scatter is reduced to 0.1. Further
discussion on the differences between FIREWORKS and SIMPLE fluxes and
derived parameters can be found in Appendix \ref{scatter}.\\

 \figp
\figq

\section{Evolution of the Specific Star Formation Rate}\label{ssfr}
In a recent paper (Damen et al. 2009a), we showed how the specific
star formation rate (SFR per unit mass, sSFR) evolves with
redshift. These findings were based on a bright sub-sample of a
preliminary version of the SIMPLE catalog, in which total fluxes were
crudely determined by applying an aperture correction to the color
fluxes. In this section, we briefly revisit the results of Damen et
al. (2009a) and show if and how they change, using the final version
of the catalog. For details on the derivation of star formation rates
and masses, see Damen et al. (2009a).\\
 \begin{deluxetable*}{lccc}
    \centering  \tablewidth{0pt}
    \tablecolumns{4}
    \tablecaption{Specific star formation rates in mass and redshift
      bins\label{ssfr.tab}
     }
     \tablehead{
      & \multicolumn{3}{c}{sSFR ($10^{-9}$ yr$^{-1}$)}\\
       \cline{2-4}
       \colhead{$z$}  & \colhead{$10 <$ log(\Mstar/\msol)$ < 10.5$} &
       \colhead{$10.5 < $log(\Mstar/\msol)$ < 11$} &
       \colhead{log(\Mstar/\msol)$ > 11$} \\
     }
     \startdata
     $0.2$ .....   & $ 0.11 \pm 0.02 $  &  $ 0.053 \pm 0.015 $  &  $     --             $  \\
     $0.4$ .....   & $ 0.23 \pm 0.02 $  &  $ 0.081 \pm 0.016 $  &  $     0.025 \pm 0.009  $  \\
     $0.6$ .....   & $    --         $  &  $ 0.19  \pm 0.02  $  &  $     0.046 \pm 0.007  $  \\
     $0.8$ .....   & $    --         $  &  $ 0.27  \pm 0.02  $  &  $     0.067 \pm 0.016  $  \\
     $1.0$ .....   & $    --         $  &  $   --            $  &  $     0.12  \pm 0.02   $  \\
     $1.2$ .....   & $    --         $  &  $   --            $  &  $     0.13  \pm 0.02   $  \\
     $1.4$ .....   & $    --         $  &  $   --            $  &  $     0.35  \pm 0.05   $  \\
     $1.6$ .....   & $    --         $  &  $   --            $  &  $     0.42  \pm 0.05   $  \\
     $1.8$ .....   & $    --         $  &  $   --            $  &  $     0.58  \pm 0.10   $  \\
     \enddata    \\                                                  
   \end{deluxetable*}                                         
For this analysis, we created a sub-sample of our catalog out to $z =
1.8$. We selected all sources with $ (\mone + \mtwo)/2 < 21.2$, a limit
that is chosen so that 95\% of the selected sources has an S/N $>$ 5 in
the K-band. From this sub-sample, we excluded all stars and all X-ray
detections, since they are likely AGNs. The final sample contains
3391 sources. Figure \ref{ssfrz} shows how the mean sSFR evolves with
redshift in different mass bins, the mean values are also given in
Table \ref{ssfr.tab}. It agrees very well with the corresponding
figure in Damen et al. (2009a), and all conclusions remain the
same. The sSFR increases with redshift for all mass bins and the slope
($dlog(sSFR)/dz$) does not seem to be a strong function of mass (see
also Damen et al. (2009b)). To quantify this, we fitted the sSFR with
$(1+z)^n$ over the redshift range where we are complete with respect
to mass. The value of the slope $n$ is 5.1$\pm$0.6 and 4.6$\pm$0.3 for
galaxies with masses $10.5 < log(M/\Msol) \le 11$ and $log(M/\Msol) >
11$, respectively. These numbers are consistent within 1-$\sigma$ with
results based on the FIREWORKS catalog over the same redshift range
(3.8$\pm$0.8 ($z < 0.8$), and 4.9$\pm$0.9 ($z < 1.8$), for both mass
bins, respectively; Damen et al. 2009b).\\

However, the number of galaxies in each bin has changed with the new
version of the catalog and this influences the fraction of quiescent
galaxies at the highest mass bin. We define a galaxy to be quiescent
when its sSFR is smaller than one-third of the inverse of the Hubble
time at its redshift ($sSFR(z) < 1/(3*t_H(z)$)). In Damen et
al. (2009a) the fraction of quiescent galaxies decreased with redshift
out to 19\%$\pm$9\% at $z$ = 1.8. Figure \ref{frac} shows the updated
version of this fraction. The old numbers are represented by dashed
lines. The slope is less steep and the fraction of quiescent galaxies
at $z$ = 1.8 is higher, 30\%$\pm7$\%. Although these numbers are
consistent within 1-$\sigma$, we investigate the cause of this change
and find that it can be explained by the increased total fluxes, which
change the masses and redshifts. The new value is in better agreement
with recent estimates Kriek et al. (2008) (36\% $\pm$ 9\%) and
I. Labb\'e et al. (in preparation) (35\% $\pm$ 7\%).\\
We have checked whether these results are robust against blending. We
redetermined mean sSFRs for two different samples, removing all
sources that (1) were flagged as blended by SExtractor and (2) we
consider blended by our own criterion. In the latter case, the mean
sSFRs change by less than 5\%, in no preferred direction and the
fraction of quiescent galaxies does not change. When all sources that
were flagged as blended by SExtractor are removed, less than 10\% of
the sources remain in each mass bin. Whereas the resulting mean sSFRs
can differ up to $\sim$40\% from the original values, they are
scattered around the mean sSFRs that are based on all sources. Hence,
the global trends stay remarkably intact and the fact that our sample
contains blended sources has no impact on the results. 

\section{Summary}\label{summary}
The {\it Spitzer} IRAC/MUSYC Public Legacy Survey in the Extended
Chandra Deep Field South (SIMPLE) consists of deep IRAC observations
(1-1.5 hr per pointing) covering the $\sim$1600 arcmin$^2$ area
surrounding the GOODS CDF-South. This region of the sky has extensive
supporting data, with deep observations from the X-rays to the thermal
infrared. We describe in detail the reduction of the IRAC observations
and the treatment of the main artifacts, such as column pulldown,
muxbleed, and muxstriping. The final SIMPLE IRAC mosaics were
complemented with 10\arcmin \, $\times$ 12\arcmin\, GOODS-IRAC images in
the center and are available online.\\
We also present a 13-band, IRAC-detected catalog based on the SIMPLE
images and existing public optical and NIR data of the MUSYC
project. The wavelength bands that are covered are $UBVRIz'JHK$ and
the four IRAC bands at 3.6, 4.5, 5.8, and 8.0 \um. The 5-$\sigma$ IRAC
depths are 23.8, 23.6, 21.9, and 21.7 for [3.6], [4.5], [5.8], and
[8.0], respectively.\\
The current catalog is an updated version of the one used in Damen et
al. (2009a). We have revisited our results in that work and found that
the conclusions stay mainly the same. Investigating the evolution of
the star formation rate we confirmed that the sSFR increases with
redshift in all mass bins and that the rate of increase
($dlog(sSFR)/dz$) does not seem to be a strong function of mass. This
is in agreement with previous work by Zheng et al. (2007) and Damen et
al. (2009b). \\
However, the redshift range over which we have determined the slope of
the sSFR is small and differs per mass bin due to incompleteness at
the low-mass end. We can use the deeper FIREWORKS catalog to
investigate the possible (lack of) evolution of $n$ with mass out to
higher redshift ($z$ = 1.5) in the three mass bins of
Fig.~\ref{ssfrz}. The values for the FIREWORKS $n$ are consistent with
the SIMPLE $n$ within 1-$\sigma$, although the number statistics in
the highest mass bin are low (on average eight sources per redshift
bin). We can conclude that the logarithmic increase of the sSFR with
redshift is at least not a strong function of mass.\\
We investigated the fraction of massive galaxies that show suppressed
star formation and found that at $z \sim 1.8$, 30\%$\pm7$\% of the
massive galaxies (\Mstar $> 10^{11}$ \Msol) have $sSFR < 1/(3 t_H)$,
which is our criterion for quenched star formation. This is consistent
within 1-$\sigma$ with the 19\%$\pm$9\% from Damen et al. (2009a) and is
in better agreement with values from Kriek et al. (2008) and
I. Labb\'e et al. (in preparation), which are 36\% $\pm$ 9\% and 35\%
$\pm$ 7\%, respectively.\\

\acknowledgments
We thank the referee for the detailed comments that helped us improve
the paper significantly.
We also thank Leonidas Moustakas for useful discussion and help with the
removal of artifacts from the SIMPLE data. This research was supported
by grants from the Netherlands Foundation for Research (NWO), and the
Leids Kerkhoven-Bosscha Fonds. Support from National Science
Foundation grant NSF CAREER AST-0449678 is gratefully acknowledged.

{\it Facilities:} \facility{Spitzer (IRAC, MIPS)}

\appendix

\section{Flux Apertures}\label{aper.app}
\figaa
When performing photometry we use SExtractor's AUTO aperture since it
is more robust than for instance the ISOCOR aperture, which depends
more sensitively on the depth of the image. In addition, it allows an
easy comparison with other catalogs such as the MUSYC and FIREWORKS
catalogs, which are both based on AUTO apertures. In
Fig.~\ref{fluxapers}, we show the effect different apertures have on 
the comparison between our catalog and the MUSYC catalog. As expected,
the AUTO fluxes give the best agreement. The cause of the offset at
the bright end of the panel showing the AUTO fluxes is discussed in
Section \ref{comp_musyc}.

\section{Confusion}\label{conf.app}
While building the SIMPLE catalog, we treated blended (or confused)
sources very conservatively and only identified the sources that most
severely suffered from blending. We were not able to simply use the
quality flags SExtractor provided, since those identified 60\% of all
sources as blended. Performing photometry on these "blended" sources
in a way commonly used for blended sources, exacerbated the
disagreement with other catalogs (see Section \ref{phot}). In
addition, it was not possible to model blended sources using a deep
source map at lower wavelength, since our $K$-band data were not deep
enough (see Section \ref{mips}). The effect blending has on photometry
is clear in, e.g., the FIREWORKS catalog, where blended sources were
identified by their SExtractor flags and take up 12\% of the
sample. Figure \ref{kblend} shows the comparison between the total
$K-$band magnitude of SIMPLE and FIREWORKS. Blended sources in the
FIREWORKS catalog are shown in red and form a distinct plume of
scattered sources. Since the plume contains only blended sources, we
removed these sources from all further analysis, since their
photometry must be inaccurate (i.e., Figures \ref{simple_fw}-\ref{fw_other}). Unfortunately, we could not apply
this trick to the SIMPLE catalog. In Section \ref{detphot}, we
identified the sources that suffer from severe blending. We have not
indicated them in Fig. \ref{kblend}, since they do not fill a specific
locus, but instead are spread out evenly over the whole figure. It is,
therefore, not possible to quantify the effect blended sources have on
our photometry and derived parameters. 
\figab
\figac
\figad

\section{Scatter between FIREWORKS and SIMPLE}\label{scatter}
In the comparison of the photometric and derived properties of the
SIMPLE and FIREWORKS catalogs, we observed a large scatter. In
Fig. \ref{mips_scatter}, we show the comparison between MIPS fluxes. The mean
values of the difference are indicated by the red line and are printed
in red in the lower right corner. Error bars represent the standard
deviation in each bin and are printed in red in the lower right
corner. The FIREWORKS MIPS fluxes have been determined based on a $K$-band
image with high spatial resolution. On the other hand, the SIMPLE
fluxes were determined using our IRAC imaging as a reference (see
Section \ref{mips}). The IRAC data are deep, but have a PSF which is much
larger, leading to more confusion. This causes the difference in MIPS
fluxes, which are relatively modest (mean absolute deviation of 10\% at
the bright end). \\
In Section \ref{derived}, we stated that the scatter in mass was not
caused by photometric redshift errors. This can be inferred from
Fig. \ref{zphot_mass}, which shows the difference in masses from
FIREWORKS and SIMPLE against spectroscopic ({\it left}) and
photometric ({\it right}) redshift. Despite the disappearance of a few
dramatic outliers, it is not clear that the scatter is much reduced
when using spectroscopic redshifts only.\\

\clearpage


\begin{thebibliography}{}
\bibitem[Alexander et al. (2008)]{alex} Alexander, D., et al. 2008,
  \aj, 135, 1968
\bibitem[Arnouts et al. (2001)]{arn01} Arnouts, S., et al. 2001, A\&A,
  379, 740
\bibitem[Ashby et al. (2009)]{ash} Ashby, L.~N. et al. 2009, \apj,
  701, 428
\bibitem[Barger et al. (1999)]{bar} Barger, A.~J., Cowie, L.~L.,
  Smail, I., Ivison, R.~J., Blain, A.~W., Kneib, J.-P. 1999, \aj, 117,
  2656
\bibitem[Bell et al. (2005)]{be05} Bell, E.~F., et al., 2005, \apj, 625, 23
\bibitem[Bertin \& Arnouts (1996)]{ba} Bertin, E. \& Arnouts, S. 1996,
  A\&AS, 117, 393
\bibitem[Bolzonella et al.(2000)]{bol} Bolzonella, M., Miralles, J.-M., Pell\o, R. 2000, A\&A, 363, 476
\bibitem[Brammer et al. (2008)]{bramm} Brammer, G.~B., van Dokkum,
  P.~G., Coppi, P. 2008, \apj, 686, 1503
\bibitem[Bruzual \& Charlot (2003)]{bc03} Bruzual, G., \& Charlot,
S. 2003, \mnras, 2, 344, 1000
\bibitem[Cardamone et al. (2008)]{card}	Cardamone, C.~N., et al. 2008, \apj, 680, 130
\bibitem[Cimatti et al. (2002)]{cim} Cimatti, A., et al. 2002, A\&A,
  392, 395 
\bibitem[Cimatti et al. (2004)]{cim2} Cimatti, A., et al. 2004,
  Nature, 430, 184 
\bibitem[Cohen et al. (2003)]{co} Cohen, M., Megeath, S.~T.,
  Hammersley, P.~L., Mart\'in-Luis, F., Stauffer, J. 2003, \aj 125,
  2645
\bibitem[Croom et al. (2001)]{cr} Croom, S.~M., Smith, R.~J., Boyle,
  B.~J., Shanks, T., Loaring, N.~S., Miller, L., Lewis, I.~J.  2001,
  \mnras, 322, 29
\bibitem[Daddi et al. (2000)]{dad1} Daddi, E., Cimatti, A., Pozzetti,
  L., Hoekstra, H., R\"ottgering, H.~J.~A., Renzini, A., Zamorani, G.,
  Mannucci, F. 2000, A\&A, 361, 535
\bibitem[Daddi et al. (2005)]{dad} Daddi, E., et al. 2005, \apj, 626,
  680
\bibitem[Daddi et al. (2007)]{dad7} Daddi, E., et al. 2007, \apj, 670,
  173
\bibitem[Dale \& Helou (2002)]{dh02} Dale, D.~A. \& Helou, G. 2002, \apj, 576, 159
\bibitem[Damen et al. (2009a)]{mcd} Damen, M., Labb\'e, I., Franx, M,
  van Dokkum, P.~G., Taylor, E.~N., Gawiser, E.~J. 2009a, \apj, 690,
  937
\bibitem[Damen et al. (2009b)]{mcd2} Damen, M., F\"orster Schreiber,
  N.~M., Franx, M., Labb\'e, I., Toft, S., van Dokkum, P.~G., Wuyts,
  S. 2009b, \apj, 705, 617
\bibitem[Dickinson et al. (2003)]{dic} Dickinson, M, et al. in The Mass
  of Galaxies at Low and High Redshift: Proc. of the European
  Southern Observatory and Universit\"ats-Sternwarte M\"unchen
  Workshop, ESO Astrophysics Symposia, Venice, Italy, 2001 October 24-26, ed. R. Bender \& A. Renzini (Berlin: Springer), 324
\bibitem[Doherty et al. (2005)]{doh} Doherty, M., Bunker, A.~J.,
  Ellis, R.~S., McCarthy, P.~J.  2005, \mnras, 361, 525
\bibitem[Donley et al. (2008)]{don} Donley, J.~L., Rieke, G.~H.,
  P\'erez-Gonz\'alez, P.~G., Barro, G. 2008, \apj, 687, 111	
\bibitem[Elston, Rieke \& Rieke (1988)]{riek} Elston, R., Rieke,
  G.~H., \& Rieke, M.~J. 1988, \apj, 331, 77
\bibitem[Erben et al. (2005)]{erbe} Erben T., et al. 2005, Astron. Nachr., 326,
  432
\bibitem[Fazio et al. (2004)]{faz} Fazio, G.~G., et al. 2004, \apjs,
  154, 10
\bibitem[Franx et al. (2003)]{franx} Franx, M. et al. 2003, \apj, 587,
  79
\bibitem[Gawiser et al. (2006)]{gaw} Gawiser, E., et al. 2006, \apjs,
  162, 1
\bibitem[Giacconi et al. (2002)]{giac} Giacconi, R., et al. 2002
  \apjs, 139, 369
\bibitem[Giavalisco et al. (2004)]{gia} Giavalisco, M., \& The GOODS
  Team 2004, ApJ, 600, L93 
\bibitem[Hildebrandt et al. (2006)]{hil} Hildebrandt, H., et al. 2006,
  A\&A, 452, 1121
\bibitem[Kriek et al. (2008)]{kriek08} Kriek, M., et al. 2008, \apj,
  677, 219
\bibitem[Kroupa, P. (2001)]{krop} Kroupa, P. 2001, \mnras, 322, 231
\bibitem[Labb\'e et al. (2003)]{lab03} Labb\'e, I. et al. 2003, \aj,
  125, 1107
\bibitem[Labb\'e et al. (2006)]{lab01} Labb\'e, I., Bouwens, R.,
  Illingworth, G,~D., Franx, M. 2006, \apj, 649, 67
\bibitem[Lacy et al. (2005)]{lac} Lacy, M., Canalizo, G., Rawlings,
  S., Sajina, A., Storrie-Lombardi, L., Armus, L., Marleau, F.~R.,
  Muzzin, A. 2005, Mem. Soc. Astron. Ital., 76, 154
\bibitem[Lai et al. (2008)] Lai, K., et al. 2008, \apj, 674, 70
\bibitem[Lehmer et al. (2005)]{lehm} Lehmer et al. 2005, \apjs, 161,
  21
\bibitem[le F\'evre et al. (2004)]{fev} le F\`evre, O., et al. 2004,
  A\&A, 428, 1043
\bibitem[Lonsdale et al. (2003)]{lon} Lonsdale, C.~J. et al. 2003,
  PASP, 115, 897
\bibitem[Luo et al. (2008)]{luo} Luo, B., et al. 2008, \apjs, 179, 19
\bibitem[Luo et al. (2010)]{luo10} Luo, B., et al.  2010, \apjs, 187, 560
\bibitem[Magnelli et al. (2009)]{mag} Magnelli, B., Elbaz, D., Chary,
  R.~R., Dickinson, M., Le Borgne, D., Frayer, D.~T., Willmer,
  C.~N.~A. 2009, A\&A, 496, 57	
\bibitem[Martin et al. (2005)]{mar} Martin, D.~C., et al. 2005, \apj,
  619, 1
\bibitem[Mignoli et al. (2005)]{mign} Mignoli, M., et al. 2005, A\&A,
  437, 883
\bibitem[Miller et al. (2008)]{mill} Miller, N.~A., Formalont, E.~B.,
  Kellermann, K.~I., Mainieri, V., Norman, C., Padovani, P., Rosati,
  P., Tozzi, P. 2008, \apjs, 179, 114
\bibitem[McCarthy et al. (2001)]{mc} McCarthy, P.~J., et al. 2001,
  \apj, 560, 131
\bibitem[Monet et al. (2003)]{mon} Monet, D.~G. et al. 2003, \aj, 125,
  984
\bibitem[Papovich et al.(2006)]{pap06} Papovich, C., et al. 2006,
  \apj, 640, 92
\bibitem[Popesso et al. (2009)]{pop} Popesso, P., et al. 2009, A\&A,
  494, 443
\bibitem[Quadri et al. (2007)]{qua} Quadri, R., et al. 2007, \aj, 134,
  1103
\bibitem[Ravikumar et al. (2006)]{ravi} Ravikumar, C. et al. 2007,
  A\&A, 465, 1099
\bibitem[Reach et al. (2005)]{rea} Reach, W.~T., et al. 2005, PASP,
  117, 978
\bibitem[Rix et al. (2004)]{rix} Rix, H.-W, et al. 2004, \apjs, 152,
  163
\bibitem[Rudnick et al. (2003)]{rud} Rudnick, G., et al. 2003, \apj,
  599, 847
\bibitem[Sanders et al. (2007)]{san} Sanders, D.~B., et al. 2007,
  \apjs, 172, 86
\bibitem[Spinrad et al. (1997)]{spin} Spinrad, H., Dey, A., Stern, D.,
  Dunlop, J., Peacock, J., Jimenez, R., Windhorst, R. 1997, \apj, 484,
  581
\bibitem[Steidel et al. (1996)]{steid96} Steidel, C.~C., Giavalisco,
  M., Dickinson, M., Adelberger, K,~L. 1996, \aj, 112, 352
\bibitem[Steidel et al. (1999)]{steid99} Steidel, C.~C., Adelberger,
  K.~L., Giavalisco, M., Dickinson, M., Pettini, M. 1999, \apj, 519, 1
\bibitem[Stern et al. (2005)]{ster} Stern, D., et al. 2005, \apj, 631,
  163
\bibitem[Strolger et al. (2004)]{strol} Strolger, L.-G., 2004, \apj,
  613, 200
\bibitem[Szokoly et al. (2004)]{szok} Szokoly, G.~P., et al.  2004,
  \apjs, 155, 271
bibitem[Taylor et al. (2009)]{enta} Taylor, E.~N., et al. 2009a, \apj,
  694, 1171
\bibitem[Taylor et al. (2009)]{entb} Taylor, E.~N., et al. 2009b, \apjs,
  183, 295
\bibitem[Treister et al. (2009)]{trea} Treister, E., et al. 2009a, \apj,
  693, 1713
\bibitem[Treister et al. (2009)]{treb} Treister, E., et al. 2009b,
  \apj, 706, 535
\bibitem[Vanzella et al. (2008)]{vanz08} Vanzella, E., et al. 2008,
  A\&A, 478, 83 
\bibitem[van der Wel et al. (2004)]{wel1} van der Wel, A., Franx, M.,
  van Dokkum, P.~G., Rix, H.-W., Illingworth, G.~D., Rosati, P. 2005,
  \apj, 631, 145	
\bibitem[van der Wel et al. (2004)]{wel2} van der Wel, A., Franx, M.,
  van Dokkum, P.~G., Rix, H.-W. 2004, \apj, 601, 5
\bibitem[van Dokkum et al. (2006)]{vdok} van Dokkum, P.~G., et
  al. 2006 \apj, 638, 59
\bibitem[Wolf et al. (2004)]{wolf} Wolf, C., et al. 2004, A\&A, 421,
  913
\bibitem[Wuyts et al. (2008)]{wuyts} Wuyts, S., Labb\'e, I.,
  F\"orster-Schreiber, N.~M., Franx, M., Rudnick, G., Brammer,
  G.~B., van Dokkum, P.~G. 2008, \apj, 682, 985
\bibitem[Zheng et al. (2007)]{zheng} Zheng, X.~Z., Bell, E.~F.,
  Papovich, C., Wolf, C., Meisenheimer, K., Rix, H.-W., Rieke, G.~H.,
  Somerville, R. 2007, \apjl, 661, 41
\end{thebibliography}
 \end{document}